%% file: HFH-for-elastic-lattices-preprint.tex
\documentclass{article}
 \usepackage[english]{ babel}
\usepackage{amsmath,amssymb}
\usepackage{graphicx}
\usepackage{caption}
\usepackage[font=footnotesize]{subcaption}
\usepackage{xfrac}
\usepackage[margin=3cm]{geometry}

\usepackage{tikz}
\usetikzlibrary{decorations.pathmorphing}
\usetikzlibrary{decorations.markings}
\usetikzlibrary{shadows}
\usetikzlibrary{calc}
\usepackage{pgfplots}

\usepackage[abs]{overpic}

\DeclareMathAlphabet\mathbit
    \encodingdefault\rmdefault\bfdefault\itdefault
\DeclareOldFontCommand{\bi}{\normalfont\bfseries\itshape}{\mathbit}

\renewcommand{\vec}[1]{{\mathbf{#1}}}
\newcommand{\vect}[1]{{\boldsymbol{#1}}}

\DeclareMathOperator{\diag}{diag}
\renewcommand{\epsilon}{\varepsilon}
\renewcommand{\phi}{\varphi}

\usepackage{sidecap}

\title{High frequency homogenisation for elastic lattices}

\author{D.J. Colquitt\footnote{Corresponding author: $\langle$ d.colquitt@imperial.ac.uk$\rangle$, $\langle$ danielcolquitt@icloud.com $\rangle$}, R.V. Craster \& M. Makwana
\\
Department of Mathematics, Imperial College London,\\
South Kensington, London, SW7 2AZ, UK}

\begin{document}

\maketitle

\begin{abstract}
A complete methodology, based on a two-scale asymptotic approach, that enables the homogenisation of elastic lattices at non-zero frequencies is developed.
Elastic lattices are distinguished from scalar lattices in that two or more types of coupled waves exist, even at low frequencies.
Such a theory enables the determination of effective material properties at both low and high frequencies.
The theoretical framework is developed for the propagation of waves through lattices of arbitrary geometry and dimension.
The asymptotic approach provides a method through which the dispersive properties of lattices at frequencies near standing waves can be described;
the theory accurately describes both the dispersion curves and the response of the lattice near the edges of the Brillouin zone.
The leading order solution is expressed as a product between the standing wave solution and long-scale envelope functions that are eigensolutions of the homogenised partial differential equation.
The general theory is supplemented by a pair of illustrative examples for two archetypal classes of two-dimensional elastic lattices.
The efficiency of the asymptotic approach in accurately describing several interesting phenomena is demonstrated, including dynamic anisotropy and Dirac cones.
\end{abstract}

\section{Introduction}

The mechanical behaviour of discrete structures is of interest in a
wide variety of physical settings, from plant and animal
tissues~\cite{gibson2010cellular}, to foams~\cite{cantat2013foams},
metamaterials for cloaking~\cite{colquitt2013making} as well as the
more familiar area of structural mechanics with applications such as sandwich plates and truss structures~\cite{wicks2001optimal}.
Much in the same way as for elastic continua, mechanical problems involving discrete structures can be broadly categorised into two kinds: \emph{scalar problems} and \emph{vector problems}.
For continuous media the material response is characterised, for
scalar problems, by a single governing partial differential equation.
For example, the shear deformation of a thin elastic membrane is governed by the Helmholtz equation with only a single degree of freedom.
Similarly, in scalar problems for discrete media, the response of the structure is governed by a single scalar difference equation with only one degree of freedom.
An example would be the out-of-plane deformation of a planar array of masses and springs.
In vector problems, the material response of the structure is
characterised by a system of governing equations, partial differential
equations in the case of continua (Navier's elastic equations for a
vector displacement) and a vector system of difference equations for
the case of discrete structures.
The analysis of discrete vector problems is far more challenging than scalar problems in a similar sense to that in continua: the elastic Lam\'{e} system is significantly more challenging than similar problems for systems governed by the Helmholtz operator. 

In terms of their static response, so-called cellular solids have been extensively studied in order to determine effective material properties (see, for example,~\cite{gibson1982mechanics,gibson1982mechanics-a,gibson2010cellular,christensen2000mechanics}).
The dispersive properties of mechanical lattices have been
analysed~\cite{martinsson2003vibrations,phani2006wave,pichard2012two} leading to many
interesting phenomena, such as dynamic
anisotropy~\cite{scarpa2002wave,ayzenberg2008resonant,osharovich2010wave,colquitt2012dynamic,movchan2013resonant},
filtering, negative refraction, and focusing in structured elastic media~\cite{colquitt2011dispersion}.
Discrete lattices have also recently been used in the design of a broadband invisibility cloak~\cite{colquitt2013making}.

Nevertheless, the determination of effective material parameters at finite frequencies remains a challenging problem.
Indeed, a proper understanding of the aforementioned phenomena is incumbent on a rigorous understanding of the material response at non-zero frequencies.
However, the homogenisation techniques applied to discrete elastic media are typically restricted to the classical long-wavelength regime (see~\cite{martinsson2007homogenization,gonella2008homogenization} among others).
Notably, for square lattices, there is a large contrast between shear and compressional deformabilities, which can result in long-wavelength compressional waves inducing short-wave resonant bending modes.
This effect was studied, using the method of multiple scales, by Chesnais~\emph{et al.}~\cite{chesnais2012effects}.

In some cases, usually for scalar problems, it is clear that a direct correspondence between discrete and continuous models exists.
An example being the normal incidence of an anti-plane shear wave on a semi-infinite stack of two dissimilar media: in this case it can be shown that the dispersion curves for the stack are asymptotically equivalent to the curves corresponding to the classical one-dimensional bi-atomic chain~\cite{movchan2002photonic,movchan2002asymptotic}.
It is also easy to observe that the difference
equations for a square
lattice mass-spring model have terms corresponding to the central
difference approximation for the Laplacian. However, in
general, and particularly for vector systems, it is unclear how to
upscale from the micro-level to an effective continuum description. 

In recent years, a two-scale asymptotic procedure has been developed which has proven an effective tool in the homogenisation of structured media at non-zero frequencies.
The method, initially developed in~\cite{craster2010high}, has been successfully
applied to a range of structured media including: continua governed by
the Helmholtz equation~\cite{craster2010high}, thin elastic plates~\cite{antonakakis2014moulding} (governed
by the Biharmonic operator), discrete scalar lattices~\cite{craster2010high-a,makwana2013localised}, networks of strings~\cite{nolde2011high} and,
more recently, the vector Lam\'{e} system~\cite{antonakakis2014moulding}. The methodology
advanced in these articles uses a separation of scales to create
effective continuum macroscale equations that incorporate the
microscale structure through coefficients of integrated
quantities. Apart from \cite{antonakakis2014moulding}, for the Lam\'{e} system, all the
examples treated thus far are limited to scalar examples and one aim
here 
is to understand how one can extend this methodology to coupled systems, particularly
given their importance as models of cellular structures.

It should be noted that other dynamic homogenisation schemes exist for structured elastic continua.
In particular, there has been much interest in the use of the Willis model~\cite{milton2007modifications} to describe the dynamic response of composites.
In recent decades, there have been many papers published with the purpose of determining the effective material properties of periodic composites that can be homogenised to the Willis model (see, for example,~\cite{willis2009exact,shuvalov2011effective,nemat2011homogenization,norris2012analytical,srivastava2012overall,srivastava2014limit}).
At present, however, the above works are restricted to composites whereas the focus of the present paper is on discrete media.
Recently, Movchan \& Slepyan analysed the dynamic response of triangular elastic lattices in the vicinity of resonant frequencies~\cite{movchan2013resonant}.
Based on a local expansion of the dispersion equation, Movchan \& Slepyan examine the behaviour of the lattice Green's function in the neighbourhood of stationary points of the dispersion surfaces.

In the present paper, we develop and implement a finite frequency homogenisation procedure in order to obtain effective continuum properties for elastic lattices governed by vectorial difference equations.
It is emphasised that this procedure is not restricted to regimes where the wavelength is much longer than the short-scale of the lattice.
The finite frequency procedure is based on the original two-scale approach~\cite{craster2010high} and extends the earlier work done for scalar lattice problems~\cite{craster2010high-a,makwana2013localised}.
The new procedure provides a general methodology for analysing any discrete periodic structure in $\mathbb{R}^d$.
Notably, the methodology incorporates the scalar theory introduced in earlier works~\cite{craster2010high-a,makwana2013localised} as a special case.

The paper is structured as follows.
The general theory and asymptotic procedure is introduced in \S\ref{sec:general-theory}.
The framework is described for lattices in arbitrary dimensions and can be applied to any class of lattice where waves propagate.
Following the general theory, the framework is applied to two archetypal lattices in \S\ref{sec:2D-lattices}: a triangular lattice (of the truss type) and a square lattice (of the frame type).
Of particular interest, are the degeneracies that can occur for lattices of the frame type; these degeneracies are analysed in detail in \S\ref{sec:square}.
Also of interest is the existence of a Dirac point in the triangular lattice and the associated linear dispersion effects.
Numerical illustrations are provided in order to illustrate the efficacy of the high frequency asymptotic procedure.
The paper is finalised with some concluding remarks in \S\ref{sec:conclusion}.

\section{General theory}
\label{sec:general-theory}

Before proceeding to the homogenisation technique, it is convenient to briefly introduce some notation.
In~\cite{martinsson2003vibrations}, Martinsson \& Movchan introduced a convenient framework in which to study the dispersion properties of discrete structures; we follow a similar approach in this section.
Consider a regular array of particles in $\mathbb{R}^d$, where $d=1,2,3,\ldots$.
Each particle in the lattice is labelled by the multi-index $\vec{m} = (m_1,\ldots,m_d)\in\mathbb{Z}^d$ and a scalar $n\in\mathbb{N}$.
The multi-index $\vec{m}$ refers to the unit cell in which the particle is located, whereas the scalar $n$ distinguishes between different particles in the same unit cell.
Introducing the direct lattice vectors $\vec{t}_i$ ($i=1,\ldots,d$), the position of each particle in the lattice is then $\vec{x}(\vec{m},n) = \mathsf{T}\vec{m} + \vec{x}(\vec{0},n)$, where $\mathsf{T} = [\vec{t}_1,\ldots,\vec{t}_d]$.
The governing equations for the time-harmonic motion of particle $(\vec{m},n)$ have the form
\begin{equation}
\omega^2\mathsf{M}(\vec{m},n)\vec{u}(\vec{m},n) = \sum_{(\vec{p},q)\in\mathcal{N}(\vec{m},n)}\mathsf{C}(\vec{p},q)\vec{u}(\vec{m}+\vec{p},q),
\label{eq:eom}
\end{equation}
where $\mathcal{N}(\vec{m},n)$ is the set of particles $(\vec{m}+\vec{p},q)$ connected to node $(\vec{m},n)$, typically this will be the set of nearest neighbours such that $\mathcal{N}(\vec{m},n) = \{(\vec{p},q)\;:\; |\vec{x}(\vec{m} + \vec{p},q) - \vec{x}(\vec{m},n)| \leq \ell\}$, where $\ell$ is the bond length.
Is is emphasised that $(\vec{m},n) \in \mathcal{N}(\vec{m},n)$.
The matrix $\mathsf{C}(\vec{p},q)$ is the stiffness matrix of the link connecting nodes $(\vec{m}+\vec{p},q)$ and $(\vec{m},n)$, the diagonal matrix $\mathsf{M}(\vec{m},n)$ describes the inertial properties of node $(\vec{m},n)$, $\vec{u}(\vec{m},n)$ is the vector of generalised displacement at node $(\vec{m},n)$, and $\omega$ is the radian frequency.
Applying the discrete Fourier transform
\begin{equation}
\vec{u}^\mathrm{F}(\vec{k},n) = \sum_{\vec{m}\in\mathbb{Z}^d} \vec{u}(\vec{m},n)e^{-i\vec{k}\cdot\vec{x}(\vec{m},n)},
\end{equation}
to the equation of motion~\eqref{eq:eom} yields
\begin{equation}
\sum_{(\vec{p},q)\in\mathcal{N}(\vec{m},n)}\left[\mathsf{C}(\vec{p},q)e^{-i\vec{x}(\vec{p})\cdot\vec{k}} -  \omega^2\mathsf{M}(n)\delta_{n,q}\right]\vec{u}^\mathrm{F}(\vec{k},q) = \vec{0},
\end{equation}
whence the dispersion equation for a perfect lattice is immediately obtained as
\begin{equation}
\det\left[\varsigma(\vec{k}) - \omega^2\mathsf{M}\right] = 0,
\end{equation}
where the matrix $\varsigma$ is partitioned as
\[
\varsigma_{nq} = \sum_{\vec{p}\in\mathcal{N}(\vec{m},n)}\left[\mathsf{C}(\vec{p},q)e^{-i\vec{x}(\vec{p})\cdot\vec{k}}\right]\quad\text{and}\quad \mathsf{M} = \mathsf{M}(n)\delta_{n,q}.
\]
It has been shown in~\cite{martinsson2003vibrations} that the matrix $\varsigma$ is Hermitian.

\subsection{The asymptotic theory}

The approached used here is that of the method of multiple scales which has already been applied to a plethora of physical problems in the setting of high frequency homogenisation~\cite{craster2010high,craster2010high-a}.
Two scales are introduced: the short-scale discrete variable $\vec{m}$, and the long-scale continuous variable $\vect{\eta} = \epsilon\mathsf{T}\vec{m}$.
It is assumed that the small parameter $0 < \epsilon \ll 1$ characterises the short-scale of the lattice.
For example, if the lattice is formed by an $N\times N$ grid of particles, where $N\gg1$, then $\epsilon = 1/N$.
The displacement is then considered as a function of two independent vector-valued variables and a single scalar variable: $\vec{u}(\vec{m},n) = \vec{u}(\vec{m},\vect{\eta},n)$.
In these two-scales the equations of motion~\eqref{eq:eom} are then written
\begin{equation}
\omega^2\mathsf{M}(\vec{m},n)\vec{u}(\vec{m},\vect{\eta},n) = \sum_{(\vec{p},q)\in\mathcal{N}(\vec{m},n)}\mathsf{C}(\vec{p},q)\vec{u}(\vec{m}+\vec{p},\vect{\eta}+\epsilon\mathsf{T}\vec{p},q).
\end{equation}
Expanding the displacement in $\vect{\eta}$ about $\vec{p} = \vect{0}$ yields
\begin{multline}
\omega^2\mathsf{M}(\vec{m},n)\vec{u}(\vec{m},\vect{\eta},n) = \\
\sum_{(\vec{p},q)\in\mathcal{N}(\vec{m},n)}\mathsf{C}(\vec{p},q)\left\{
\vec{u}(\vec{m}+\vec{p},\vect{\eta},q) + \epsilon\left[\mathsf{T}\vec{p}\cdot\nabla \right]\vec{u}(\vec{m}+\vec{p},\vect{\eta},q)
\vphantom{\frac{\epsilon^2}{2}}\right. \\
\left. + \frac{\epsilon^2}{2}\mathsf{T}\vec{p}\cdot\left[\mathsf{T}\vec{p}\cdot\nabla\left(\nabla\vec{u}(\vec{m}+\vec{p},\vect{\eta},q)\right)\right]
\right\} + \mathcal{O}(\epsilon^3),
\end{multline}
where $\nabla$ acts on the continuous long-scale variable $\vect{\eta}$.
In essence, the high frequency homogenisation approach involves perturbing away from standing wave frequencies.
The change of phase across the unit cell is described by the function $e^{-i\vec{x}(\vec{p})\cdot\vec{k}}$, where $\vec{k}$ is the point of interest in Fourier space.
Imposing this phase shift only on the short-scale variable yields
\begin{multline}
\omega^2\mathsf{M}(\vec{m},n)\vec{u}(\vec{m},\vect{\eta},n) = \\
\sum_{(\vec{p},q)\in\mathcal{N}(\vec{m},n)}\mathsf{C}(\vec{p},q)e^{-i\vec{x}(\vec{p})\cdot\vec{k}}\left\{
\vec{u}(\vec{m}+\vec{p},\vect{\eta},q) + \epsilon\left[\mathsf{T}\vec{p}\cdot\nabla \right]\vec{u}(\vec{m}+\vec{p},\vect{\eta},q)
\vphantom{\frac{\epsilon^2}{2}}\right. \\
\left. + \frac{\epsilon^2}{2}\mathsf{T}\vec{p}\cdot\left[\mathsf{T}\vec{p}\cdot\nabla\left(\nabla\vec{u}(\vec{m}+\vec{p},\vect{\eta},q)\right)\right]
\right\} + \mathcal{O}(\epsilon^3).
\end{multline}
The expanded system is conveniently written as
\begin{equation}
\omega^2\mathsf{M}(\vec{m})\vec{u}(\vec{m},\vect{\eta}) = \left[\sigma_0(\vec{k}) + \epsilon\sigma_1(\vec{k}) + \epsilon^2\sigma_2(\vec{k})\right]\vec{u}(\vec{m},\vect{\eta})
\label{eq:expanded-one}
\end{equation}
where the higher order terms have been suppressed, but are considered understood, and the matrices $\sigma_i$ are partitioned as follows
\begin{align}
\left[\sigma_0(\vec{k})\right]_{nq} & =
\sum_{\vec{p}\in\mathcal{N}(\vec{m},n)}\mathsf{C}(\vec{p},q)e^{-i\vec{x}(\vec{p})\cdot\vec{k}}, \\
\label{eq:sigma1}
\left[\sigma_1(\vec{k})\right]_{nq} & =
\sum_{\vec{p}\in\mathcal{N}(\vec{m},n)}\mathsf{C}(\vec{p},q)e^{-i\vec{x}(\vec{p})\cdot\vec{k}}\left[\mathsf{T}\vec{p}\cdot\nabla \right], \\
\left[\sigma_2(\vec{k})\right]_{nq} & =
\sum_{\vec{p}\in\mathcal{N}(\vec{m},n)}\mathsf{C}(\vec{p},q)e^{-i\vec{x}(\vec{p})\cdot\vec{k}}\;
\frac{1}{2}\mathsf{T}\vec{p}\cdot\left[\mathsf{T}\vec{p}\cdot\nabla\left(\nabla\right)\right],
\end{align}
together with
\[
\vec{u}(\vec{m},\vect{\eta}) = \begin{bmatrix}
\vec{u}(\vec{m},\vect{\eta},1) \\
\vdots\\
\vec{u}(\vec{m},\vect{\eta},P)
\end{bmatrix},\quad\text{and}\quad
\mathsf{M}(\vec{m}) = \diag[\mathsf{M}(\vec{m},1),\ldots,\mathsf{M}(\vec{m},P)].
\]
Here, $P$ is the number of particles in the elementary cell.
The form of~\eqref{eq:expanded-one} suggests the following ansatz for
the displacement and frequency squared 
\begin{equation}
\vec{u}(\vec{m},\vect{\eta}) = \sum_{n=0}^\infty \epsilon^n\vec{u}^{(n)}(\vec{m},\vect{\eta}),\quad
\omega^2 = \sum_{n=0}^\infty \epsilon^n\omega_n^2.
\end{equation}
Substituting this ansatz into~\eqref{eq:expanded-one} yields a hierarchy of equations in ascending orders of $\epsilon$, the first three of which are
\begin{align}
\label{eq:leading-order}
\left[\omega_0^2\mathsf{M} - \sigma_{0}(\vec{k})\right]\vec{u}^{(0)}(\vect{\eta}) & = 0, \\
\label{eq:first-order}
\left[\omega_0^2\mathsf{M} - \sigma_{0}(\vec{k})\right]\vec{u}^{(1)}(\vect{\eta}) & = \left[\sigma_{1}(\vec{k}) - \omega_1^2\mathsf{M}\right]\vec{u}^{(0)}(\vect{\eta}), \\
\label{eq:second-order}
\left[\omega_0^2\mathsf{M} - \sigma_{0}(\vec{k})\right]\vec{u}^{(2)}(\vect{\eta}) & = \left[\sigma_{1}(\vec{k}) - \omega_1^2\mathsf{M}\right]\vec{u}^{(1)}(\vect{\eta}) + \left[\sigma_{2}(\vec{k}) - \omega_2^2\mathsf{M}\right]\vec{u}^{(0)}(\vect{\eta}),
\end{align}
where the dependence of $\vec{u}$ on $\vec{m}$ has been suppressed but is considered understood.
The form of the leading order equation admits a solution with the decomposition $\vec{u}^{(0,i)} = \phi^{(0,i)}(\vect{\eta})\vec{U}^{(0,i)}(\vec{m})$, where the index $i$ enumerates the eigensolutions.
It is remarked that this decomposition holds only if the corresponding eigenvalue $\omega_{(0,i)}^2$ is simple, i.e. has unitary multiplicity.
In what follows, it will be assumed that the eigenvalue has a multiplicity of one.
For the case of eigenvalues with non-unitary multiplicities, the following analysis is subject to small technical modifications, but the same approach as used for simple eigenvalues remains valid.

The scalar function $\phi^{(0,i)}$ describes the behaviour of the lattice on the long-scale, whereas the vector $\vec{U}^{(0,i)}$ relates to the short-scale behaviour.
Whilst this may initially seem counterintuitive, that the long-scale behaviour of the vectorial system can be described by a scalar function, it is in fact the case and there is no contradiction.
Indeed for all vector problems, $\mathsf{M}^{-1}\sigma_{0}(\vec{k})$ is a square matrix of size $N>1$.
Hence, the leading order problem~\eqref{eq:leading-order} has $N>1$ eigensolutions.
Physically, this means that the long-scale response of elastic lattices governed by vectorial equations is characterised by two or more scalar functions.
The remainder of this section will be devoted to the determination of the scalar functions $\phi^{(0,i)}$.

Moving to the first order equation~\eqref{eq:first-order}, it is clear from the leading order problem~\eqref{eq:leading-order} that the bracketed term is singular and hence, according to the Fredholm alternative\footnote[2]{
One could also use the necessary and sufficient condition for the existence of solutions of the linear system $\mathsf{A}\vec{x} = \mathsf{b}$, that is, $[\mathsf{I} -\mathsf{A}\mathsf{A}^+]\vec{b} = \vec{0}$ (see~\cite{penrose1955generalized,benisrael1974generalized} among others).}
, the first order problem is solvable iff
\begin{equation}
{\vec{U}^{(0,i)}}^{\dagger}\sigma_{1}(\vec{k})\vec{U}^{(0,i)}\phi^{(0,i)}(\vect{\eta}) - \omega_{(1,i)}^2\mathsf{M}\left|\vec{U}^{(0,i)}\right|^2\phi^{(0,i)}(\vect{\eta}) = 0,
\label{eq:first-order-pde}
\end{equation}
where $(\cdot)^\dagger$ denotes the Hermitian transpose.
Usually, but not always, ${\vec{U}^{(0,i)}}^{\dagger}\sigma_{1}(\vec{k})\vec{U}^{(0,i)} = 0$ in which case $\omega_{(1,i)} = 0$ and the next-to-leading order solution is of the form
\[
\vec{u}^{(1,i)}(\vect{\eta}) = \mathsf{S}_{(0,i)}^+\sigma_1\vec{U}^{(0,i)}\phi^{(0,i)}(\vect{\eta})
+ \left[\mathsf{I} - \mathsf{S}_{(0,i)}^+\mathsf{S}_{(0,i)}\right]\vec{v}\psi(\vect{\eta}),
\]
where $\mathsf{S}_{(0,i)} = \omega_{(0,i)}^2\mathsf{M} - \sigma_0$, $\vec{v}$ is an arbitrary vector, $\psi(\vect{\eta})$ is an arbitrary scalar function of $\vect{\eta}$, $\mathsf{I}$ is the identity matrix and $(\cdot)^+$ denotes the [Moore-Penrose] pseudoinverse~\cite{penrose1955generalized,benisrael1974generalized}.
In the event that $\omega_{(1,i)}$ is non-zero,
then~\eqref{eq:first-order-pde} is a first order partial differential
equation for the envelope function $\phi^{(0,i)}(\vect{\eta})$. The
partial differential equation is posed entirely upon the long-scale
and is the homogenised equation that captures behaviour near the
standing wave frequency.

If $\omega_{(1,i)}$ vanishes (i.e. if ${\vec{U}^{(0,i)}}^{\dagger}\sigma_{1}(\vec{k})\vec{U}^{(0,i)} = 0$), then it is necessary to consider the second order problem.
Applying the same solvability criterion to~\eqref{eq:second-order} yields
\begin{multline}
{\vec{U}^{(0,i)}}^{\dagger}\sigma_{1}\mathsf{S}_{(0,i)}^+\sigma_1\vec{U}^{(0,i)}\phi^{(0,i)}(\vect{\eta})
+ {\vec{U}^{(0,i)}}^{\dagger}\sigma_{1}\left[\mathsf{I} - \mathsf{S}_{(0,i)}^+\mathsf{S}_{(0,i)}\right]\vec{v}\psi(\vect{\eta}) \\
+ {\vec{U}^{(0,i)}}^{\dagger}\sigma_{2}\vec{U}^{(0,i)}\phi^{(0,i)}(\vect{\eta}) - \omega_{(2,i)}^2\left|\vec{U}^{(0,i)}\right|^2\phi^{(0,i)}(\vect{\eta})= 0.
\label{eq:second-order-pde-part1}
\end{multline}
Since $\left[\mathsf{I} - \mathsf{S}_{(0,i)}^+\mathsf{S}_{(0,i)}\right]$ is the orthogonal projector onto the kernel of $\mathsf{S}_{(0,i)}$, 
\begin{align*}
{\vec{U}^{(0,i)}}^{\dagger}\sigma_{1}\left[\mathsf{I} - \mathsf{S}_{(0,i)}^+\mathsf{S}_{(0,i)}\right]\vec{v}\psi(\vect{\eta}) & =
{\vec{U}^{(0,i)}}^{\dagger}\sigma_{1}\vec{U}^{(0,i)}\psi(\vect{\eta}) \\
& = 0.
\end{align*}
Hence, for simple eigenvalues, the term involving the arbitrary vector $\vec{v}$ in~\eqref{eq:second-order-pde-part1} vanishes and the solvability condition reduces to
\begin{multline}
{\vec{U}^{(0,i)}}^{\dagger}\sigma_{1}\mathsf{S}_{(0,i)}^+\sigma_1\vec{U}^{(0,i)}\phi^{(0,i)}(\vect{\eta})
+ {\vec{U}^{(0,i)}}^{\dagger}\sigma_{2}\vec{U}^{(0,i)}\phi^{(0,i)}(\vect{\eta})\\
 - \omega_{(2,i)}^2\left|\vec{U}^{(0,i)}\right|^2\phi^{(0,i)}(\vect{\eta})= 0.
\label{eq:second-order-pde}
\end{multline}
Provided that $\omega_{(2,i)}$ is non-zero, equation~\eqref{eq:second-order-pde} yields a second order partial differential equation of the scalar envelope function $\phi^{(0,i)}(\vect{\eta})$.
In this case, the dispersion curves will be locally quadratic. The
partial differential equation is, again, posed entirely upon the
long-scale and creates the effective homogenised equation that
represents the behaviour of the medium. 
In the case of vanishing $\omega_{(2,i)}$, one would again repeat the previous steps moving to higher order equations.

\subsubsection{Eigenvalues of non-unitary multiplicity}

If the eigenvalue $\omega_{(0,i)}^2$ has multiplicity $R$, then the leading order problem admits solutions of the form
\[
\vec{u}^{(0,i)}(\vect{\eta}) = \sum_{s=1}^R \phi^{(0,i,s)}(\vect{\eta})\vec{U}^{(0,i,s)}.
\]
The solvability condition for the $\mathcal{O}(\epsilon)$ problem is then
\begin{equation}
\sum_{s=1}^R\sum_{t=1}^R {\vec{U}^{(0,i,s)}}^{\dagger} \left[ \sigma_{1}(\vec{k}){\vec{U}^{(0,i,t)}}(\vect{\eta}) - \omega_{(1,i)}^2\mathsf{M}{\vec{U}^{(0,i,t)}}\right]\phi^{(0,i,s)}\phi^{(0,i,t)} = 0,
\label{eq:1st-order-solvability-multiple}
\end{equation}
where the dependence of $\phi^{(0,i,t)}$ on $\vect{\eta}$ has been suppressed.
Provided $\omega_{(1,i)}$ is non-zero,~\eqref{eq:1st-order-solvability-multiple} yields a system of $R$ equations from which the envelope functions $\phi^{(0,i,t)}$ are determined.
In this case, the dispersion curves are locally linear and such points are associated with Dirac cones.
If $\omega_{(1,i)}$ vanishes, then we proceed to higher order as before.
The solvability condition again yields a system of $R$ second order equations as in the first order problem.
Upon decoupling the system, a single partial differential equation is obtained for the envelope functions.
In the case of vanishing $\omega_{(2,i)}$, one would again repeat the previous steps moving to higher order equations.

\section{Two dimensional elastic lattices: trusses and frames}
\label{sec:2D-lattices}

The approach of high frequency homogenisation has already been successfully applied to one- and two-dimensional scalar lattices (those corresponding to the out-of-plane displacement of particles connected via massless springs).
Although the general theory presented in \S\ref{sec:general-theory} is equally applicable to one- and three- dimensional lattices, for the sake of clarity we exclusively consider two-dimensional lattices.
Two dimensional lattices allow us to demonstrate all the salient phononic features of mechanical lattices without unnecessarily obfuscating the presentation.

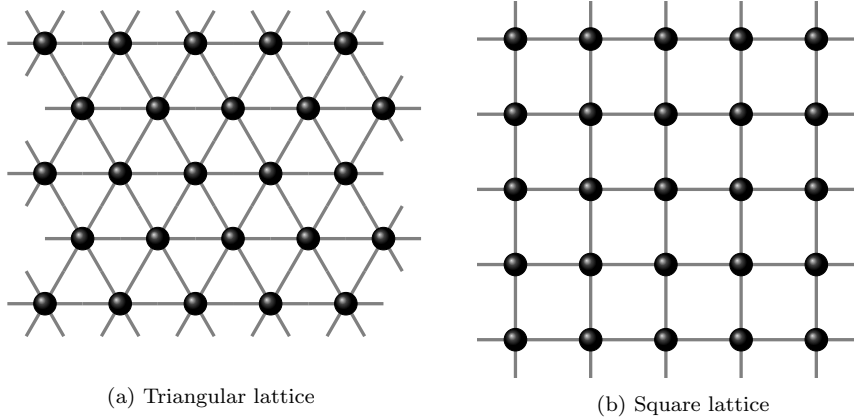
\begin{figure}
\centering
\begin{subfigure}[c]{0.35\linewidth}
\input{uniform-triangular-lattice}
\caption{
Triangular lattice}
\end{subfigure}
\qquad
\begin{subfigure}[c]{0.35\linewidth}
\input{uniform-square-lattice}
\caption{
Square lattice}
\end{subfigure}
\caption{
\label{fig:lattice-schematic}
Schematics of the two exemplar lattices considered.
Physically, the lattices can be thought of as point masses connected by thin elastic massless links.
}
\end{figure}

Physically, two-dimensional mechanical lattices can be thought of as a regular array of point masses distributed over the plane $\mathbb{R}^2$ and connected by thin elastic massless rods.
There are two fundamental types of mechanical structures: (a) \emph{truss-like}, and (b) \emph{frame-like}.
Truss-like structures are those in which the longitudinal stiffness of the lattice links dominates the problem and the flexural stiffness can be neglected.
In this case, the force exerted between two particles is generated purely by the extension and compression of the link.
The junctions between the links behave as pin-joints:
triangular lattices are an example of truss-like structures.
Conversely, frame-like structures are those in which the flexural stiffness of the links must be accounted for, otherwise the structure becomes degenerate (see, for example,~\cite{deshpande2001foam}).
Here, the links are treated as massless Euler-Bernoulli beams connecting point masses.
In the case of frame-like structures the angle at which the links meet is fixed and the junctions are endowed with a polar mass moment of inertia:
 square and honeycomb lattices are examples of frame-like structures.
Alternatively, square lattices can also be interpreted as two-dimensional arrays of cylinders connected by elastic springs~\cite{pichard2012two}.
In both cases the lattice links are assumed to be slender (i.e. the thickness of the links is assumed to be small when compared with the length); this is particularly important in the case of frames so that shear effects can be neglected.
Further details on the distinction between trusses and frames can be found in~\cite{martinsson2003vibrations,deshpande2001foam}.
In the current section we will examine the dispersive properties of uniform triangular and square lattices (see figure~\ref{fig:lattice-schematic}) within the framework of high frequency homogenisation.

\subsection{Triangular lattices}
\label{sec:triangular}

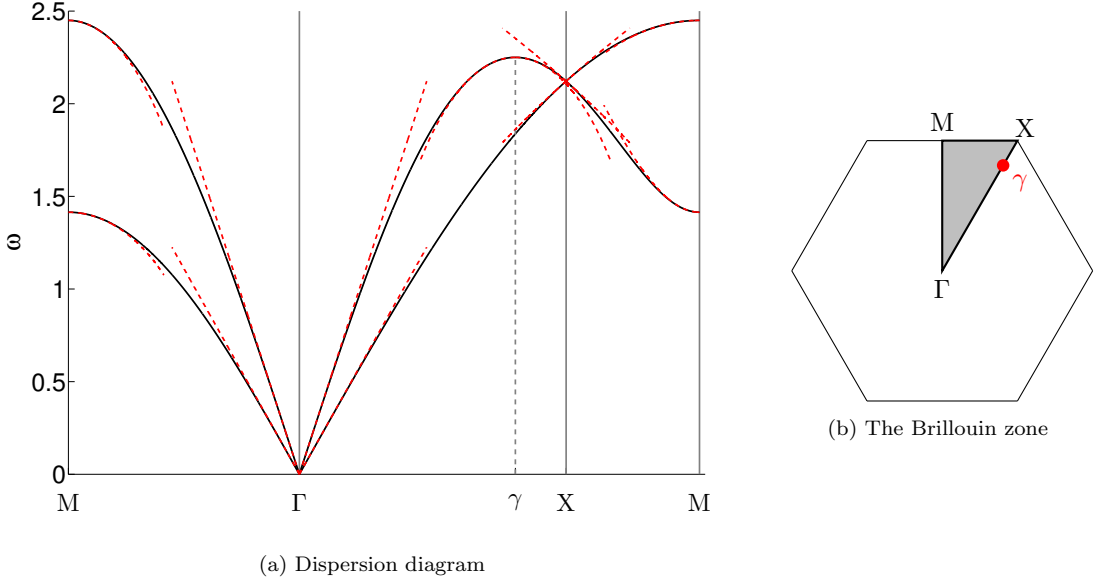
\begin{figure}
\centering
\begin{subfigure}[c]{0.7\linewidth}
\includegraphics[width=\linewidth]{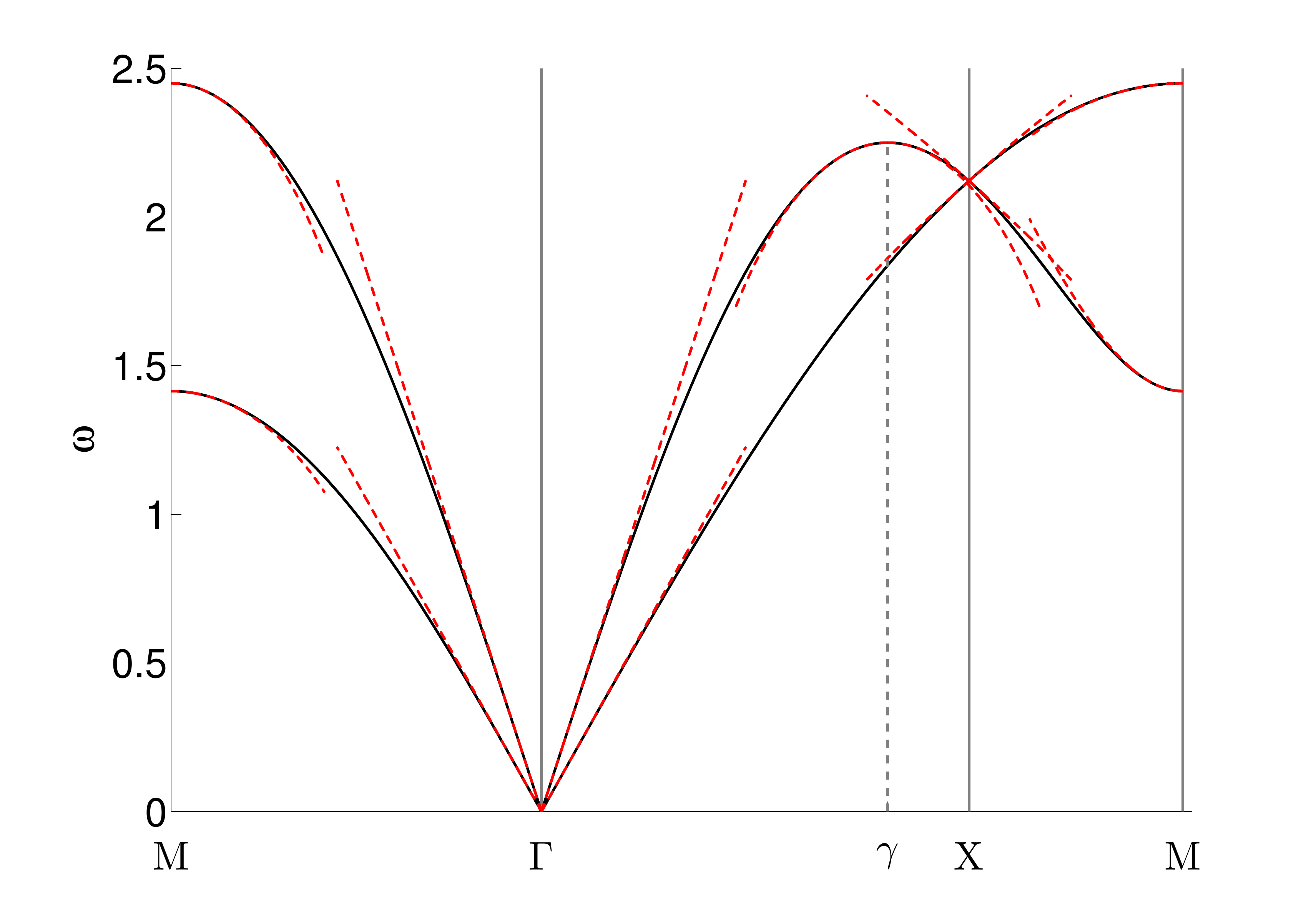}
\caption{
Dispersion diagram}
\end{subfigure}
\begin{subfigure}[c]{0.25\linewidth}
\begin{tikzpicture}[scale=1.5]
\foreach \n in {0,1,...,5} {
	\draw[rotate around = {60*\n:(0,0)}] ({-2/3},{2/sqrt(3)}) -- ({2/3},{2/sqrt(3)});
	}
\draw[thick,fill=lightgray] (0,0) -- (0,{2/sqrt(3)}) -- ({2/3},{2/sqrt(3)}) -- cycle;
\node[below] at (0,0) {$\displaystyle \Gamma$};
\node at ({2/3*1.1},{2/sqrt(3)*1.1}) {$\displaystyle \mathrm{X}$};
\node[above] at (0,{2/sqrt(3)}) {$\displaystyle \mathrm{M}$};
\draw node[fill=red,circle,scale=0.5] at (0.5398930876,0.9351222585) {};
\node[below right] at (0.5398930876,0.9351222585) {$\displaystyle\color{red}\gamma$};
\end{tikzpicture}
\caption{
The Brillouin zone}
\end{subfigure}
\caption{
\label{fig:disp-brillouin}
The Brillouin zone and dispersion curves (black solid) and finite frequency asymptotics (dashed red) for the triangular elastic lattice.
The corners of the irreducible Brillouin zone are $\Gamma=\left(0,0 \right), \mathrm{M} =\left(0,2\pi/\sqrt{3} \right)$ and $\mathrm{X}=\left(2\pi/3, 2\pi/\sqrt{3} \right)$. }
\end{figure}

Using the general methodology of \S\ref{sec:general-theory}, the [non-dimensionalised] equations of motion for the time-harmonic displacement of a uniform triangular elastic lattice are
\begin{multline}
-\omega^2\mathsf{M}(\vec{m})\vec{u}(\vec{m}) =
\mathsf{C}_1\left[\vec{u}(\vec{m}+\vec{e}_1) + \vec{u}(\vec{m}-\vec{e}_1) - 2\vec{u}(\vec{m})\right] + \\
\mathsf{C}_2\left[\vec{u}(\vec{m}+\vec{e}_2) + \vec{u}(\vec{m}-\vec{e}_2) - 2\vec{u}(\vec{m})\right] +\\
\mathsf{C}_3\left[\vec{u}(\vec{m}+\vec{e}_1-\vec{e}_2) + \vec{u}(\vec{m}-\vec{e}_1+\vec{e}_2) - 2\vec{u}(\vec{m})\right],
\label{eq:triangular-eom}
\end{multline}
where $\vec{e}_i = [\delta_{1i},\delta_{2i}]^\mathrm{T}$ are enumeration vectors and the stiffnesses of the lattice links are
\begin{equation}
\mathsf{C}_1 = \begin{pmatrix}
1 & 0 \\
0 & 0
\end{pmatrix}, \quad
\mathsf{C}_2 = \frac{1}{4}\begin{pmatrix}
1 & \sqrt{3} \\
\sqrt{3} & 3
\end{pmatrix}, \quad
\mathsf{C}_2 = \frac{1}{4}\begin{pmatrix}
1 & -\sqrt{3} \\
-\sqrt{3} & 3
\end{pmatrix}.
\end{equation}
The position of each node in the lattice is given by $\vec{x}(\vec{m}) = [m_1+m_2/2,m_2\sqrt{3}/2]^\mathrm{T}$ and the inertia matrix is simply $\mathsf{M} = \diag[1,1]$.
The non-dimensional frequency is related to the physical frequency thus $\omega^2 = \tilde{\omega}^2m/\mu$, where $m$ is the particle mass, and $\mu$ and $\ell$ are the stiffness and length of the rods respectively.
The displacement vectors are also normalised by the length of the rods $\vec{u}(\vec{m}) = [u_1(\vec{m})/\ell, u_2(\vec{m})/\ell]^\mathrm{T}$, where $u_i(\vec{m})$ are the displacements along the coordinate axes.
Since the lattice is uniform, there is only a single node in the elementary cell and the scalar index $n$ is therefore omitted.
For an infinite lattice, the two branches of the dispersion surfaces are (see, for example,~\cite{movchan2013resonant})
\begin{multline}
\omega_{\pm}^2 = 1 - \cos k_1 + 2\left[ 1 - \cos\left(\frac{k_1}{2}\right)\cos\left(k_2\frac{\sqrt{3}}{2}\right)\right] \\
\pm\left\{\left[ \cos k1 - \cos\left(\frac{k_1}{2}\right)\cos\left(\frac{k_1\sqrt{3}}{2}\right)\right]^2 + 3\sin^2\left(\frac{k_1}{2}\right)\sin^2\left(k_2\frac{\sqrt{3}}{2}\right)\right\}^\frac{1}{2}
\end{multline}
There are four points of interest on the boundary of the irreducible Brillouin zone (see figure~\ref{fig:disp-brillouin}): $\Gamma$, $\mathrm{X}$, $\mathrm{M}$, and $\gamma$.
The first three points lie at the corners of the irreducible Brillouin zone, whilst the third lies along the edge $\Gamma\mathrm{X}$.
We proceed with the high frequency homogenisation procedure by introducing a small parameter $0 < \epsilon \ll 1$ and a slow variable $\vect{\eta} = \epsilon[m_1+m_2/2,m_2\sqrt{3}/2]^\mathrm{T}$.
The ansatz for the field and frequency are introduced and a hierarchy of equations in ascending orders of $\epsilon$ is obtained, as described in \S\ref{sec:general-theory}.

\begin{figure}
\centering
\begin{subfigure}[c]{0.4\linewidth}
\includegraphics[width=\linewidth]{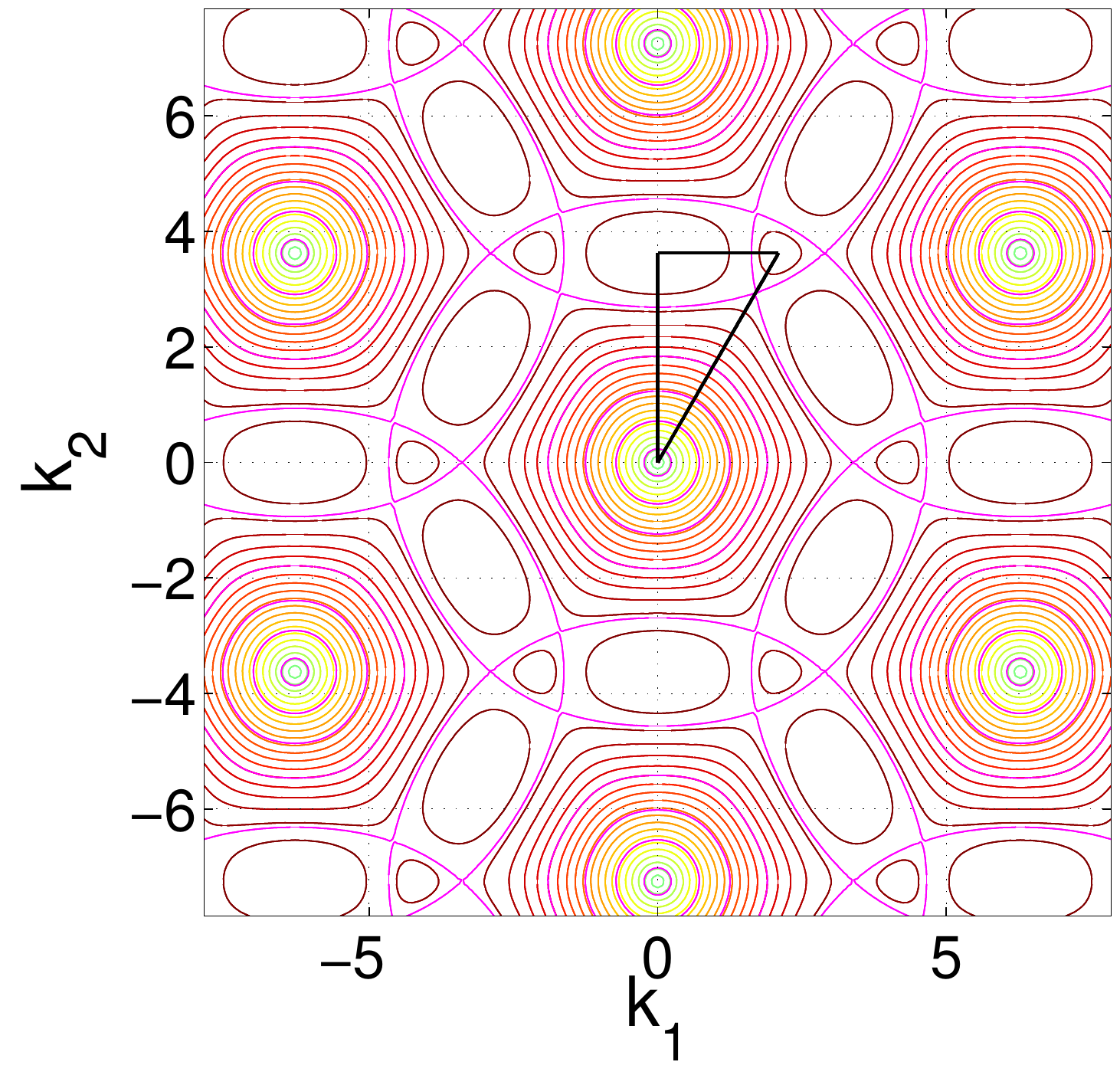}
\caption{
The upper dispersion surface $\omega_+$.}
\end{subfigure}
\qquad
\begin{subfigure}[c]{0.4\linewidth}
\includegraphics[width=\linewidth]{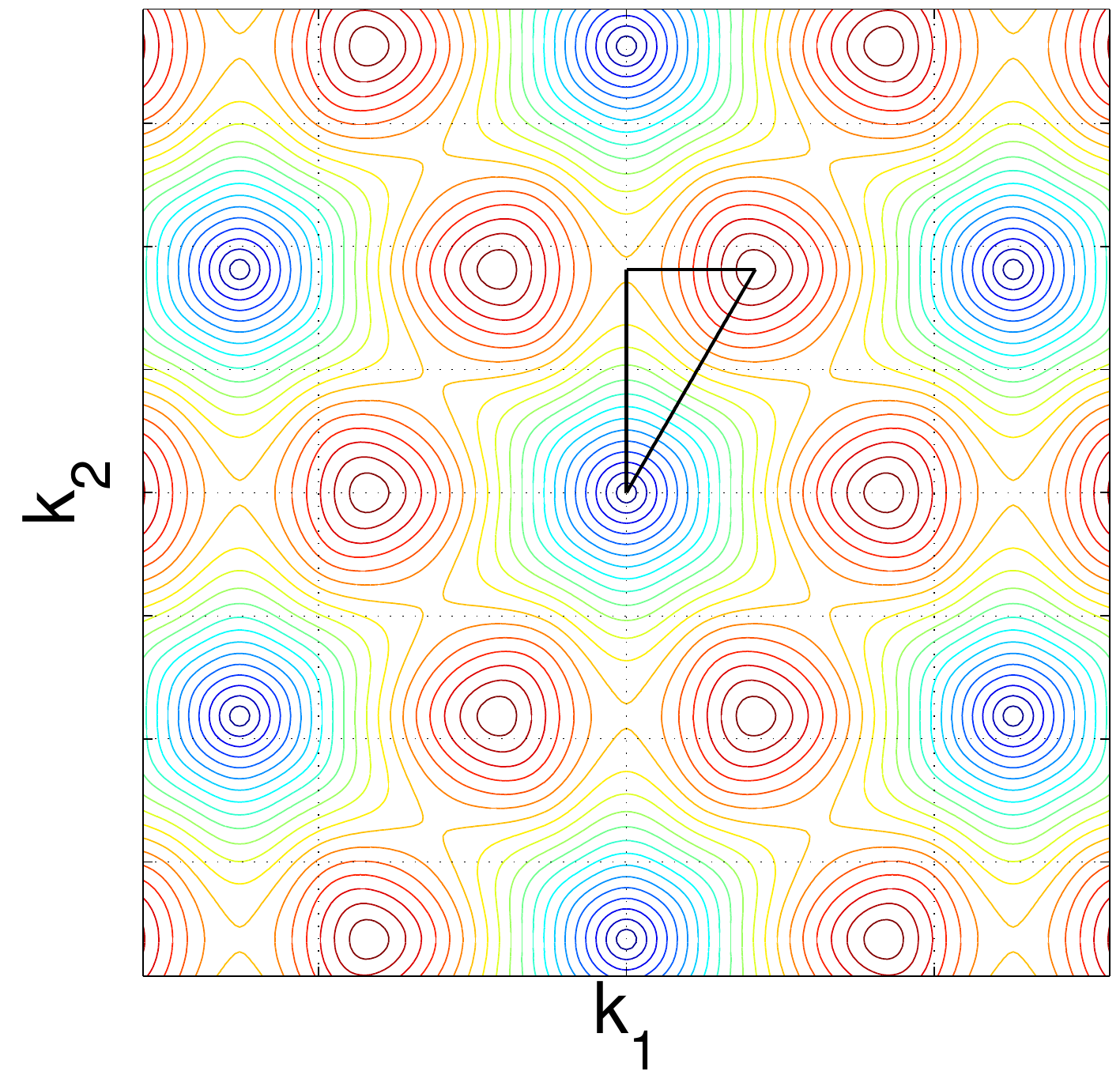}
\caption{
The lower dispersion surface $\omega_-$}
\end{subfigure}
\caption{\label{fig:slowness}
The irreducible Brillouin zone superimposed on the isofrequency curves for the triangular lattice.}
\end{figure}

\subsubsection{Numerics}
\label{sec:numerics}
In order to corroborate our asymptotic method it is convenient to have an efficient independent numerical alternative; with this motivation in mind, we truncate the infinite system to a finite system, containing $N$ masses in the direction of the lattice basis vectors. The equations of motion~\eqref{eq:triangular-eom} are reformulated into the following matrix problem
\begin{subequations}
\begin{multline}
\hspace{-0.5cm} \left(\mathsf{D}_1+m\omega^2 \mathsf{I}_N \right) \vect{U}_1 + \frac{1}{4}\left[\left(\vec{U}_1+\sqrt{3} \vec{U}_2 \right)\mathsf{D}_2+\mathsf{D}_3\left(\vec{U}_1-\sqrt{3} \vec{U}_2 \right)\mathsf{D}_3
\right. \\ \left.
+ \mathsf{D}_3^{\dagger}\left(\vec{U}_1-\sqrt{3} \vec{U}_2 \right)\mathsf{D}_3^{\dagger} \right]=\vec{0}
\end{multline}
\begin{multline}
\left(M\omega^2 -3 \right) \vec{U}_2 + \frac{\sqrt{3}}{4}\left[\left(\vec{U}_1+\sqrt{3} \vec{U}_2 \right)\mathsf{D}_2-\mathsf{D}_3\left(\vec{U}_1-\sqrt{3} \vec{U}_2 \right)\mathsf{D}_3
\right.\\ \left.
-\mathsf{D}_3^{\dagger}\left(\vec{U}_1-\sqrt{3} \vec{U}_2 \right)\mathsf{D}_3^{\dagger} \right]=\vec{0}, 
\end{multline}
\end{subequations}
where $\mathsf{D}_3$ is an $N^2$ matrix containing $1$ along a single
secondary diagonal, $\mathsf{I}_N$ is the $N^2$ identity matrix,
$\mathsf{D}_1= \mathsf{D}_2-3\mathsf{I}_N,
\mathsf{D}_2=\mathsf{D}_3+\mathsf{D}_3^{\dagger}$ and $\vec{U}_1,
\vec{U}_2$ are the displacement matrices. In order to limit spurious
reflections associated to propagating solutions, PML-like boundary
conditions~\cite{makwana2013localised} are applied around the edges of
the domain. When verifying our homogenisation method, we shall solely
deal with defect-modes derived by a specified directional
forcing. Hence the above matrix equations will be augmented with
forcing terms of the right-hand sides and then solved accordingly.

\subsubsection{At point $\mathbf{\Gamma}$: Classical, low-frequency homogenisation}
At point $\Gamma$, $\vec{k} = \vec{0}$, the phase shift across the cells is zero: all nodes move in-phase.
In this case, the leading order problem is
\begin{equation}
-\omega_{(0)}^2\vec{u}^{(0)}(\vect{\eta}) = \vect{0},
\end{equation}
whence, for non-trivial solutions, we deduce that $\omega_{(0)} = 0$ and the leading order solution admits the representation
\begin{equation}
\vec{u}^{(0)} = \sum_{s=1}^2 \phi^{(0,s)}\vec{U}^{(0,s)},
\end{equation}
where $\vec{U}^{(0,s)}$ are any constant linearly independent vectors.
Moving to the $\mathcal{O}(\epsilon)$ equation, we again find
\[
-\omega_{(1)}^2\vec{u}^{(1)}(\vect{\eta}) = \vect{0},
\]
and hence $\omega_{(1)} = 0$.
Finally, the $\mathcal{O}(\epsilon^2)$ equation is
\begin{multline}
\vect{0} = 
\mathsf{M}\omega_{(2)}^2\vec{u}^{(0)}{(\vect{\eta})} +
\left(\mathsf{C}_1 + \frac{1}{4}\mathsf{C}_2 + \frac{1}{4}\mathsf{C}_3\right)\partial^2_1\vec{u}^{(0)}(\vect{\eta})
+\frac{3}{4}\left(\mathsf{C}_2 + \mathsf{C}_3\right)\partial^2_2\vec{u}^{(0)}(\vect{\eta}) \\
+\frac{\sqrt{3}}{2}\left(\mathsf{C}_2 - \mathsf{C}_3\right)\partial_1\partial_2\vec{u}^{(0)}(\vect{\eta}),
\end{multline}
where $\partial_i = \partial/\partial\eta_i$ denotes differentiation with respect to the long-scale variables.
Forming the solvability condition and assuming, without loss of generality, that $\vec{U}^{(0,s)} = [\delta_{1s},\delta_{2s}]^\mathrm{T}$; yields the following system of coupled partial differential equations
\begin{subequations}
\begin{equation}
0 =
\omega_{(2)}^2\phi^{(0,1)}(\vect{\eta}) +
\frac{9}{8}\partial^2_1\phi^{(0,1)}(\vect{\eta})
+\frac{3}{8}\partial^2_2\phi^{(0,1)}(\vect{\eta})
+\frac{3}{4}\partial_1\partial_2\phi^{(0,2)}(\vect{\eta}),
\end{equation}
\begin{equation}
0 =
\omega_{(2)}^2\phi^{(0,2)}(\vect{\eta}) +
\frac{3}{8}\partial^2_1\phi^{(0,2)}(\vect{\eta})
+\frac{9}{8}\partial^2_2\phi^{(0,2)}(\vect{\eta})
+\frac{3}{4}\partial_1\partial_2\phi^{(0,1)}(\vect{\eta}).
\end{equation}
\end{subequations}
The system decouples as
\begin{equation}
\left[\frac{3}{8}\nabla^2 + \omega_{(2)}^2\right]\left[\frac{9}{8}\nabla^2 + \omega_{(2)}^2\right]\phi^{(0,i)}(\vect{\eta}) = 0,\qquad\text{for}\;i=1,2.
\label{eq:pointA-homogenised}
\end{equation}
For a perfect lattice, $\phi^{(0,i)}(\vect{\eta}) = \phi^{(0,i)}\exp(i\vec{k}\cdot\vect{\eta}/\epsilon)$ and we immediately obtain the dispersion branches near the origin
\begin{equation}
\omega \sim |\vec{k}|\sqrt{\frac{3}{8}},\qquad \text{and}\qquad \omega \sim |\vec{k}|\sqrt{\frac{9}{8}}.
\label{eq:tri-disp-A}
\end{equation}
Equation~\eqref{eq:pointA-homogenised} is the partial differential equation governing the envelope functions $\phi^{(0,i)}$ on the long scale.
The lattice has thus been homogenised into an effective continuum, with the long scale behaviour governed by~\eqref{eq:pointA-homogenised} and the short-scale oscillations described by the vectors $\vec{U}^{(0,i)}$.
The reader's attention is drawn to the fact that~\eqref{eq:pointA-homogenised} is written entirely on the long-scale, in terms of macroscopic variables.
As expected, the response of the lattice is isotropic and the two acoustic dispersion surfaces are linear in the low-frequency regime.
Indeed, one observes the circular isofrequency curves near the origin in figure~\ref{fig:slowness}, indicating an isotropic response in the low frequency regime.

From the low-frequency dispersion equations~\eqref{eq:tri-disp-A} one can infer the classical shear and compressional wave speeds: $c_s^2 = 3/8$ and $c_p^2 = 9/8$, from which one can further deduce the effective elastic moduli: $\lambda=\mu = \sqrt{3}/4$.
These agree with those results found by static analysis~\cite{gibson2010cellular}, energy considerations~\cite{ostoja2002lattice}, or similar analysis of the dispersion equations~\cite{phani2006wave}.
Furthermore, it is clear from equation~\eqref{eq:pointA-homogenised} that the envelope functions $\phi^{(0,i)}$ can be viewed as linear combinations of the shear and compressional potentials.

\subsubsection{Point $\mathbf{M}$: Band-edges and resonant waveforms}

Proceeding as before, at point $M$ with $\vec{k} = [0,\pi]^\mathrm{T}$, the leading order problem is
\begin{equation}
\omega_{(0)}^2\vec{u}^{(0)}(\vect{\eta}) = 4\left(\mathsf{C}_2 + \mathsf{C}_3\right)\vec{u}^{(0)}
\end{equation}
and the solvability criterion yields two simple eigenvalues $\omega_{(0,1)}^2 = 2$ and $\omega_{(0,2)}^2 = 6$.
The leading order solutions admit the decomposition $\vec{u}^{(0,i)} = \phi^{(0,i)}(\vect{\eta})\vec{U}^{(0,i)}$, for $i=1,2$ and $\vec{U}^{(0,i)} = [\delta_{1i},\delta_{2i}]^\mathrm{T}$.
Moving to the leading order problem, we once again find that $\omega_{(1,i)} = 0$ and, since $\vec{u}_{(1,i)}$ does not appear in the $\mathcal{O}(\epsilon^2)$ problem, the next to leading order solution is irrelevant.
Imposing the solvability condition on the second order problem yields the following system of uncoupled partial differential equations
\begin{subequations}
\label{eq:pointB-homogenised}
\begin{equation}
\frac{7}{8}\partial^2_1\phi^{(0,1)}(\vect{\eta}) - \frac{3}{8}\partial^2_2\phi^{(0,1)}(\vect{\eta}) + \omega_{(2,1)}^2\phi^{(0,1)}(\vect{\eta}) = 0,
\label{eq:pointB-homogenised-hyperbolic}
\end{equation}
\begin{equation}
-\frac{3}{8}\partial^2_1\phi^{(0,2)}(\vect{\eta}) - \frac{9}{8}\partial^2_2\phi^{(0,2)}(\vect{\eta}) + \omega_{(2,2)}^2\phi^{(0,2)}(\vect{\eta}) = 0.
\label{eq:pointB-homogenised-elliptic}
\end{equation}
\end{subequations}
The uncoupled system~\eqref{eq:pointB-homogenised} describes the effective continuum, or long-scale, behaviour of the lattice close to the two resonant frequencies $\omega_{(0,1)}$ and $\omega_{(0,2)}$.
The first partial differential equation~\eqref{eq:pointB-homogenised-hyperbolic} is hyperbolic and is associated with dynamic anisotropy and resonant waveforms near $\omega_{(0,1)} = \sqrt{2}$.
In this case waves propagate along the principle directions, defined by the characteristics of equation~\eqref{eq:pointB-homogenised-hyperbolic}, but decay exponentially in all other directions.
These directions of preferential propagation can also be inferred from the hexagonal and star-shaped isofrequency contours plotted in figure~\ref{fig:slowness}; see~\cite{ayzenberg2008resonant,osharovich2010wave,colquitt2012dynamic} for further details.

The second equation~\eqref{eq:pointB-homogenised-elliptic} is elliptic and is associated with the global maximum of the dispersion surfaces close to $\omega_{(0,2)} = \sqrt{2}$.
If $\omega_{(2,1)}$ is real, corresponding to a perturbation into the stop-band of the lattice, then it is clear that~\eqref{eq:pointB-homogenised-elliptic} has only evanescent solutions.
On the other hand, if $\omega_{(2,1)}$ is purely imaginary then~\eqref{eq:pointB-homogenised-elliptic} has propagating solutions.
This corresponds to a perturbation into the pass-band.
The reader's attention is drawn to the fact that~~\eqref{eq:pointB-homogenised-elliptic} is anisotropic.

\begin{figure}
\centering
\begin{subfigure}[image1]{0.45 \linewidth}
\includegraphics[width=\linewidth]{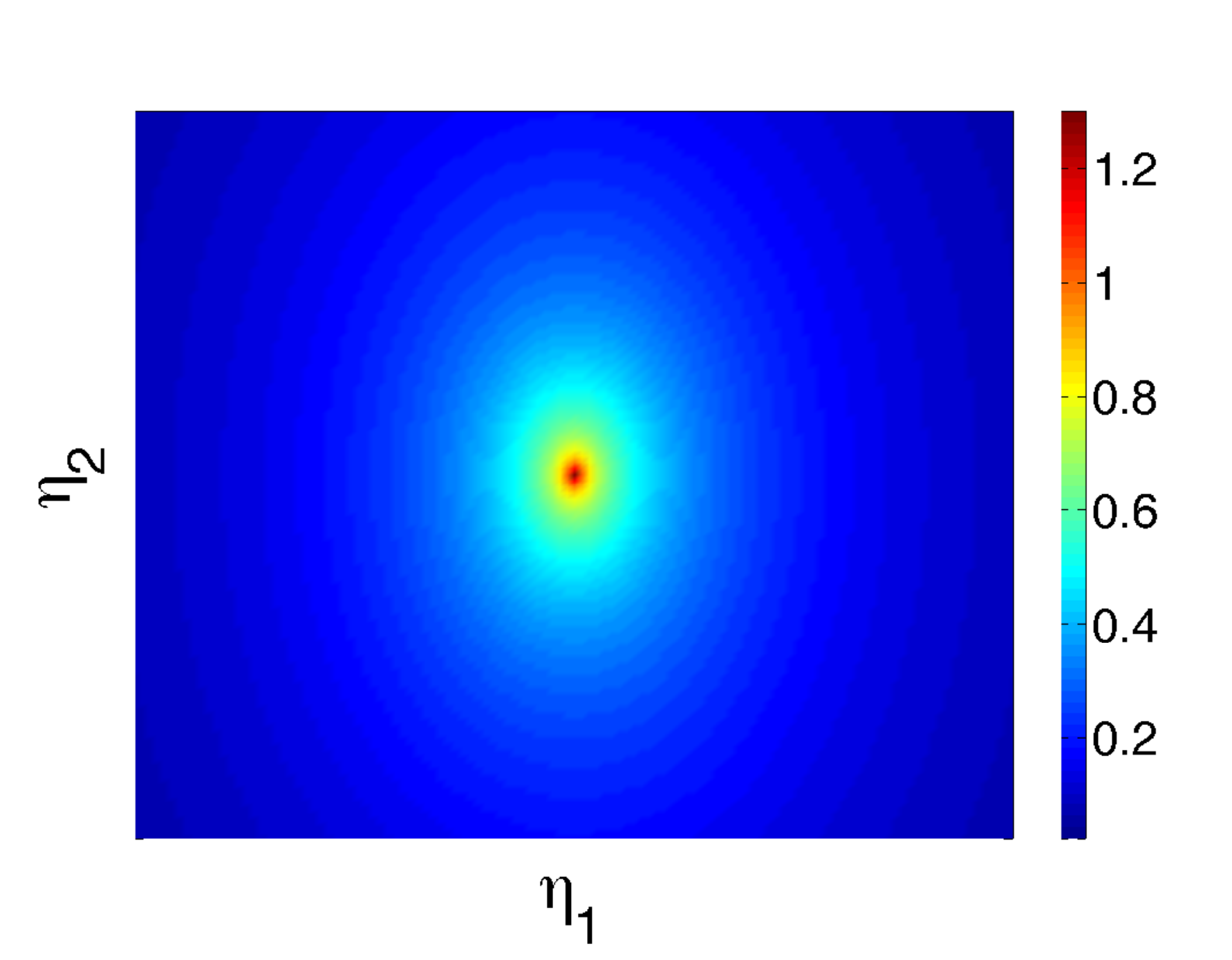}
\caption{\label{fig:pointB1_pcol}
$|\vec{u}(\vec{m})|$}
\end{subfigure}
\begin{subfigure}[image2]{0.5\linewidth}
\includegraphics[width=\linewidth]{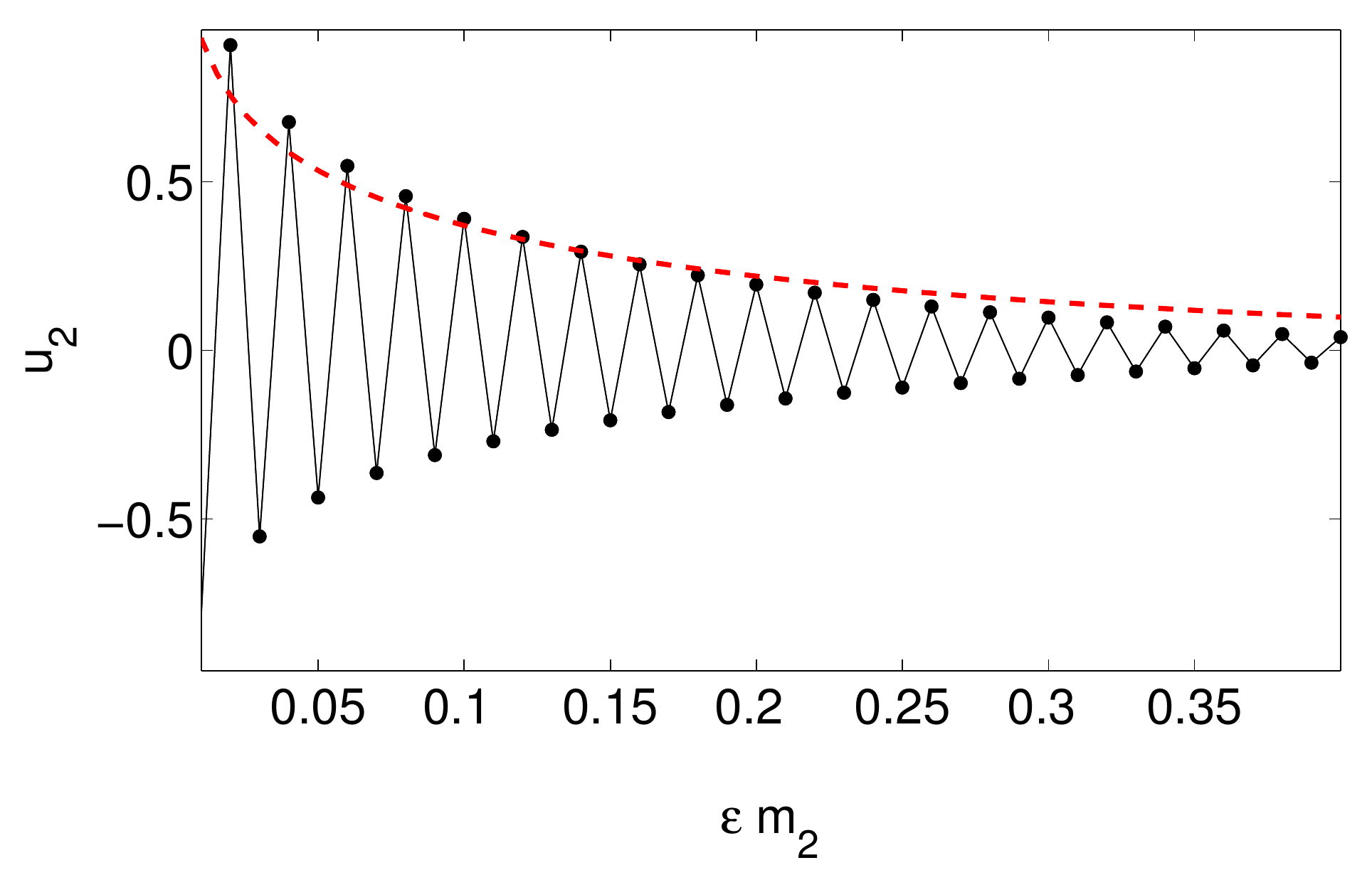}
\caption{\label{fig:pointB1_comp}
$u_2(0,\epsilon m_2)$}
\end{subfigure}
\begin{subfigure}[image3]{0.45 \linewidth}
\includegraphics[width=\linewidth]{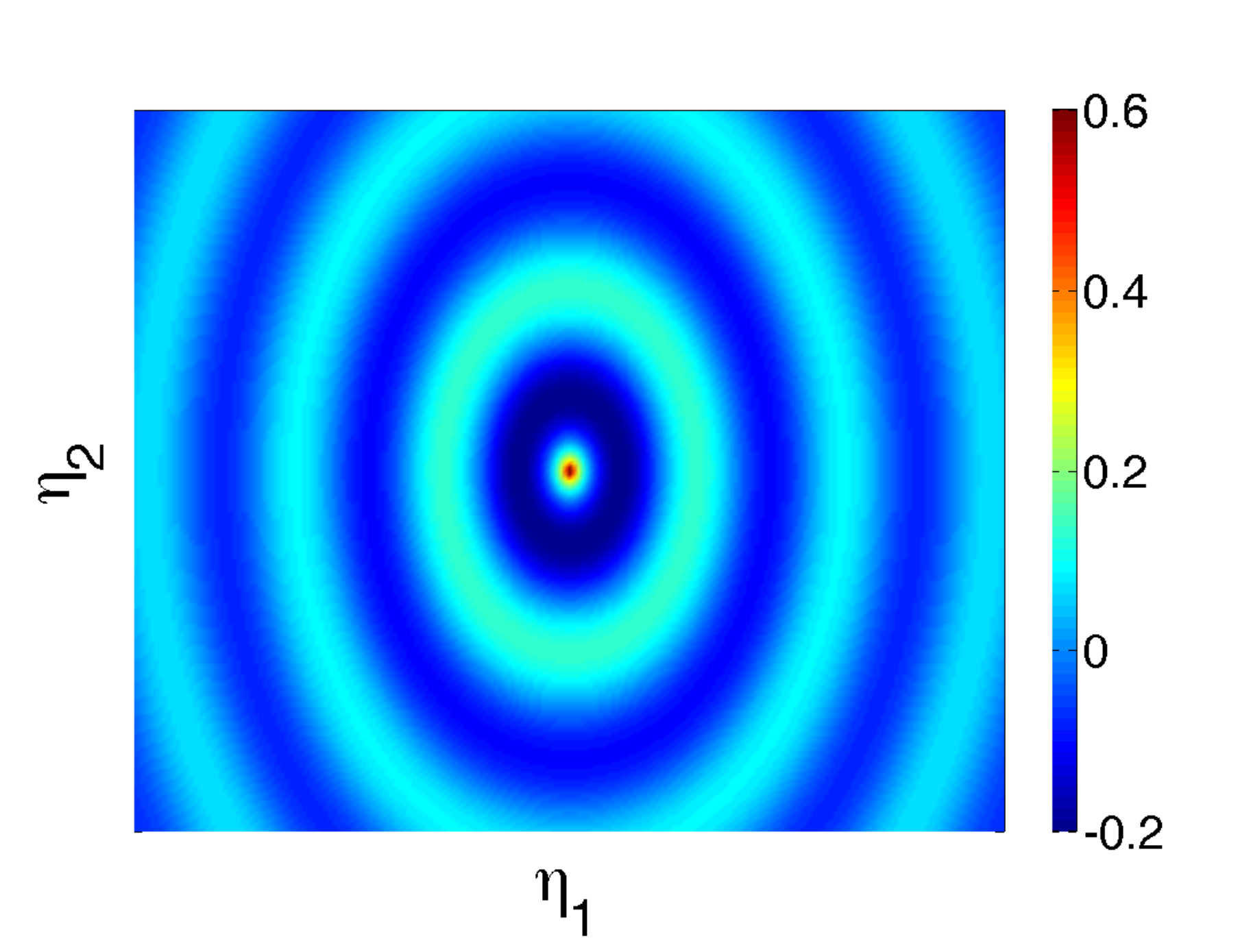}
\caption{\label{fig:pointB1_pcol2}
$u_2(\vec{m})$}
\end{subfigure}
\begin{subfigure}[image4]{0.5\linewidth}
\includegraphics[width=\linewidth]{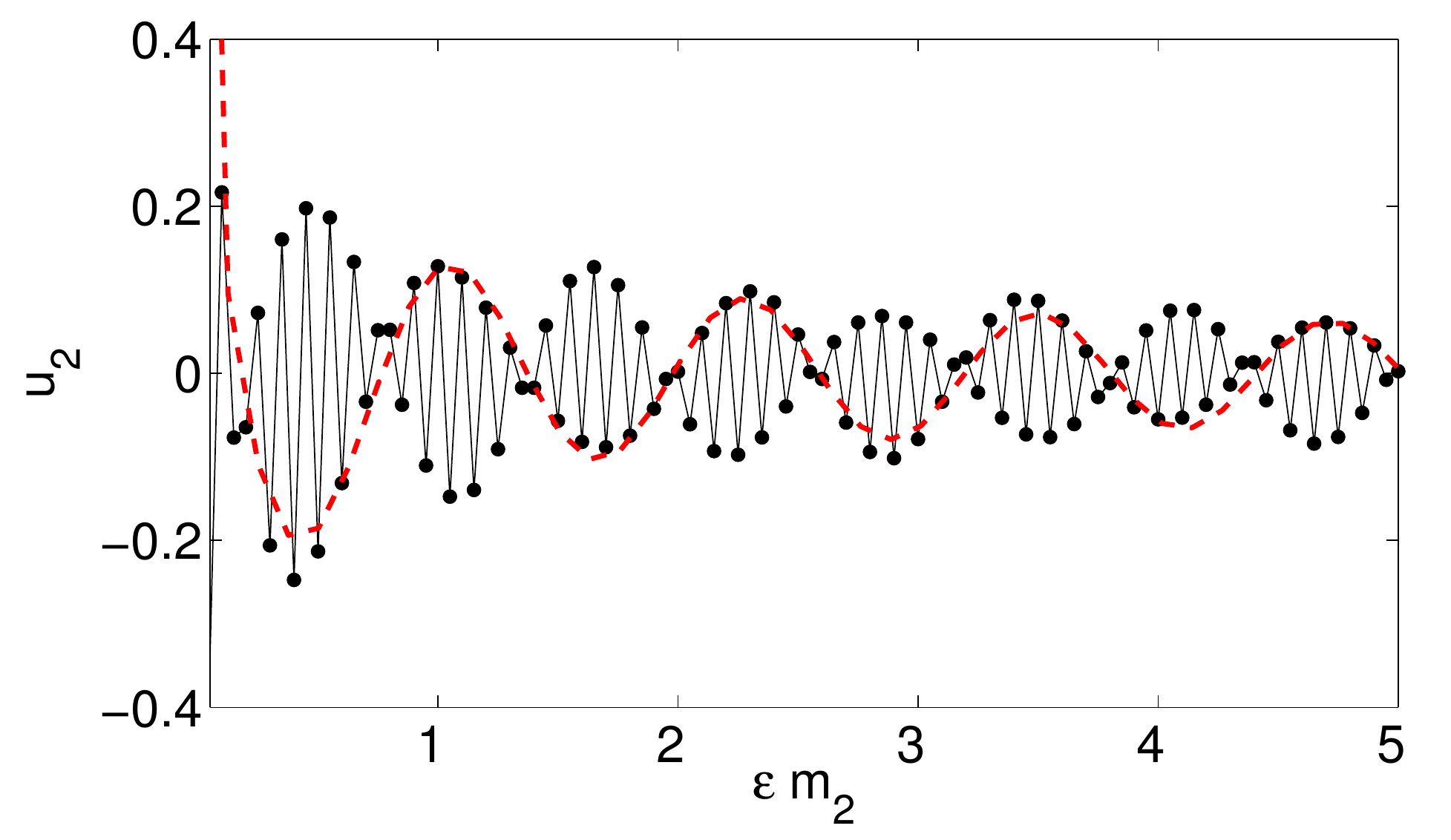}
\caption{\label{fig:pointB1_comp2}
$u_2(0,\epsilon m_2)$}
\end{subfigure}
\caption{
Panels (a) \& (b) show the evanescent solution ($\epsilon=0.01, \omega_{(2,2)}^2=5$), whilst panels (c) \& (d) show the propagating solution ($\epsilon=0.1, \omega_{(2,2)}^2=-5$).
A pseudocolour plot of $|{\bf u\left({\bf m} \right)}|$ is shown in panel (a), whilst $u_2(\vec{m})$ is plotted in panel (c).
The numerical solution (solid black) is compared with the asymptotic envelope (red dashed) for the displacement $u_2\left(0,\epsilon m_2 \right)$ in panels (b) \& (d) for the evanescent and propagating solutions respectively.}
\label{fig:pointB1_comp}
\end{figure}

We proceed with verifying the efficacy of our method, using the matrix approach outlined in section~\ref{sec:numerics}.
As we wish to analyse modes located within both the stop and pass-bands, our focus shall be on the global maximum.
An $\mathcal{O}\left(\epsilon \right)$ vertical excitation is applied to the central mass; the magnitude and direction of the forcing ensures that the term propagates to the desired second-order equation.
The resulting PDE is identical to ~\eqref{eq:pointB-homogenised-elliptic} albeit with the forcing term, $F \delta(\vect{\eta})$, located on the right-hand side.
Solving this equation gives us the leading-order envelope modulation 
\begin{equation}
\phi^{(0,2)}(\vec{\eta})=\frac{2i F}{3\sqrt{3}} H_0^{(1)} \left(i \omega_{(2,2)} \sqrt{\frac{8 M}{3}}\left[\eta_1^2+\frac{\eta_2^2}{3} \right] \right).
\label{eq:pointB-elliptic-forced-solution}
\end{equation}
A direct comparison between the asymptotics and numerics within both, the decaying and propagating regions, is shown in figure \ref{fig:pointB1_comp}.
The anisotropic nature of the oscillations is also demonstrated.

\subsubsection{At point $\mathbf{X}$: Dirac cones}
\label{sec:triangular-X}

\begin{SCfigure}
\centering
\includegraphics[width=0.5\linewidth]{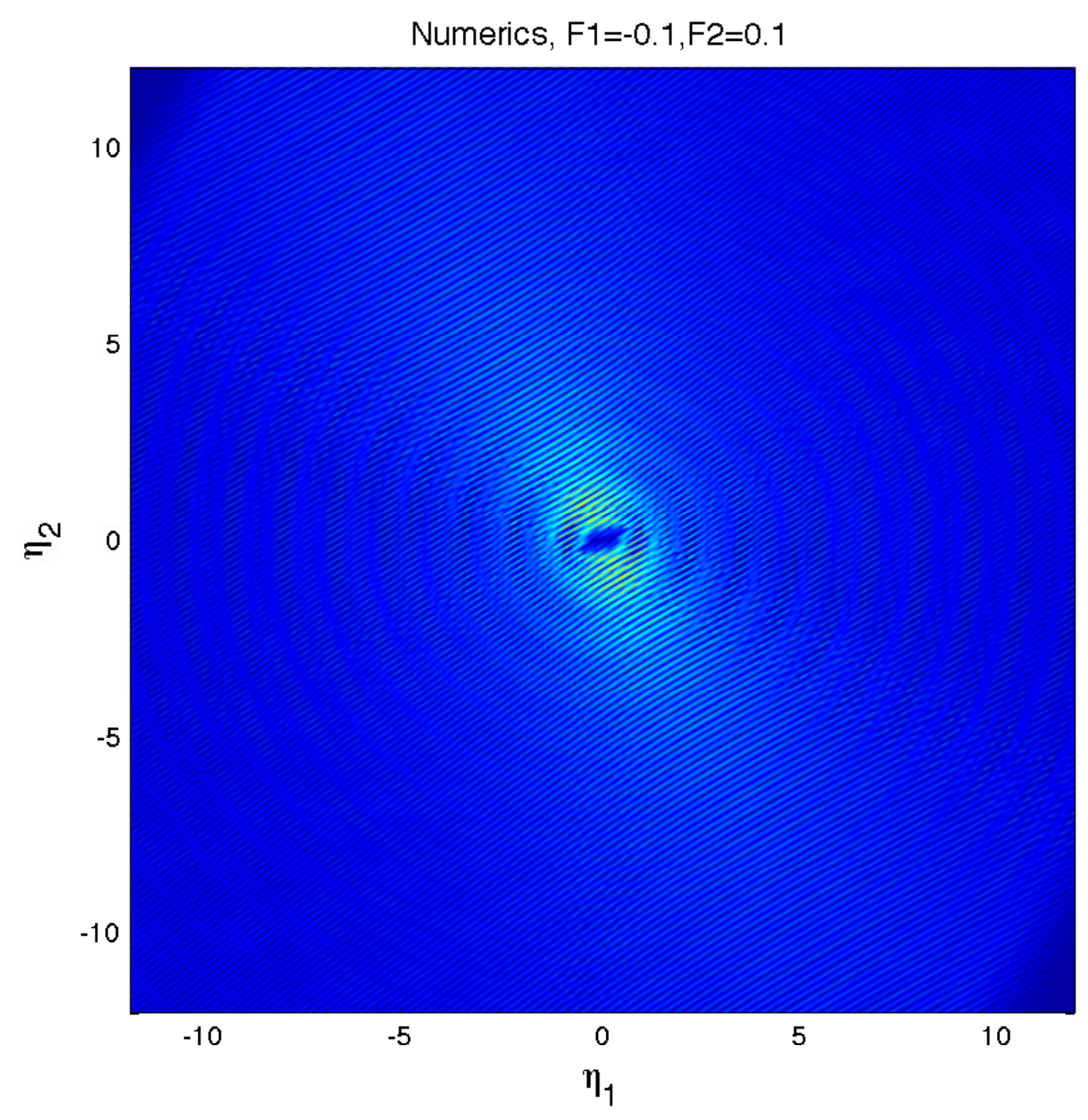}
\caption{\label{fig:dirac-cones}
The displacement field generated by forcing a small cluster of nodes at a frequency close the standing wave frequency at $X$ ($\omega = \sqrt{\omega_{(0)}^2 -  0.08}$).
The response is approximately isotropic; the small level of anisotropy can be attributed to the excited of a non-resonant mode associate with the other dispersion branch.}
\end{SCfigure}

At point $X$, with $\vec{k} = [2\pi/3,2\pi/\sqrt{3}]^\mathrm{T}$, the Bloch-Floquet quasi-periodicity envelope is $\exp[i2\pi/3(m_1 +2 n_2)]$ and the leading order problem is
\begin{equation}
\omega_{(0)}^2\vec{u}^{(0)}(\vect{\eta}) = 3\left(\mathsf{C}_1 + \mathsf{C}_2 + \mathsf{C}_3\right)\vec{u}^{(0)}(\vect{\eta}).
\end{equation}
The solvability criterion yields an eigenvalue (of multiplicity two) $\omega_{(0)}^2 = 9/2$, whence the leading order solution admits the decomposition
\[
\vec{u}^{(0)}(\vect{\eta}) = \phi^{(0,1)}(\vect{\eta})\vec{U}^{(0,1)} + \phi^{(0,2)}(\vect{\eta})\vec{U}^{(0,2)},
\]
with $\vec{U}^{(0,i)} = [\delta_{1i},\delta_{2i}]^\mathrm{T}$.
Moving to the $\mathcal{O}(\epsilon)$ problem, we find that $\omega_{(1,i)}$ does not necessarily vanish and instead we obtain the following coupled system of partial differential equations
\begin{subequations}
\begin{equation}
i \omega_{(1)}^2\phi^{(0,1)}(\vect{\eta}) = \frac{3\sqrt{3}}{4}\left[\partial_1\phi^{(0,1)}(\vect{\eta}) - \partial_2\phi^{(0,2)}(\vect{\eta})\right],
\end{equation}
\begin{equation}
i \omega_{(1)}^2\phi^{(0,2)}(\vect{\eta}) = -\frac{3\sqrt{3}}{4}\left[\partial_1\phi^{(0,2)}(\vect{\eta}) + \partial_2\phi^{(0,1)}(\vect{\eta})\right]
\end{equation}
\end{subequations}
The system then decouples into a pair of second order partial differential equations
\begin{equation}
\frac{27}{16}\nabla^2\phi^{(0,i)}(\vect{\eta}) + \omega_{(1)}^4\phi^{(0,i)}(\vect{\eta}) = 0,\qquad\text{for}\, i=1,2.
\end{equation}
In this case, the the dispersion curves are locally linear and the response is isotropic, as expected at Dirac points.

Dirac cones are of particular interest due to their dispersive properties and are the subject of significant attention in the physics community, particularly with regard to graphene~\cite{neto2009electronic}.
Dirac-like cones have also been found in other physical systems, such as platonic cystals~\cite{antonakakis2014moulding,smith2014double}, honeycomb lattices coupled with flexural plates~\cite{torrent2013elastic} and, more recently, in phononic crystals~\cite{antonakakis2014homogenisation}.
Although Dirac cones may be found in crystals with square geometry (see, for example,~\cite{antonakakis2014moulding}), hexagonal lattices are of particular interest since there is a connection between such geometries and band surfaces that give rise to Dirac cones; this connection is independent of governing equation~\cite{peleg2007conical}.

Nevertheless, it is interesting to observe that Dirac-like points also occur in discrete elastic systems.
We note that since the material parameters have been scaled out of the physical system, the Dirac point at $X$ is stable; that is, it will persist regardless of the material parameter of the lattice.

\subsubsection{At point $\mathbf{\gamma}$: A saddle point}
\label{sec:tri-gamma}

A saddle point of the dispersion surfaces occurs at $\vec{k} = \{\pi-\arccos(1/8)\}[1,\sqrt{3}]^\mathrm{T}$.
Although there appears to be a local maximum in figure~\ref{fig:disp-brillouin} it is, in fact, a saddle point.
This illustrates a key problem with simply plotting dispersion curves around a  contour following the edge of the Brillouin zone~\cite{craster2012dangers}: it is possible to misidentify, or miss entirely, stationary points on the dispersion surfaces.
Nevertheless, the nature of the stationary point is immediately apparent as a result of the homogenisation method used here; the form of the homogenised partial differential equation is related to the type of stationary point considered.
In particular, at saddle points the homogenised partial differential equation will be hyperbolic, as we have seen earlier.

\begin{figure}
\centering
\begin{subfigure}{0.32\linewidth}
\includegraphics[width=\linewidth]{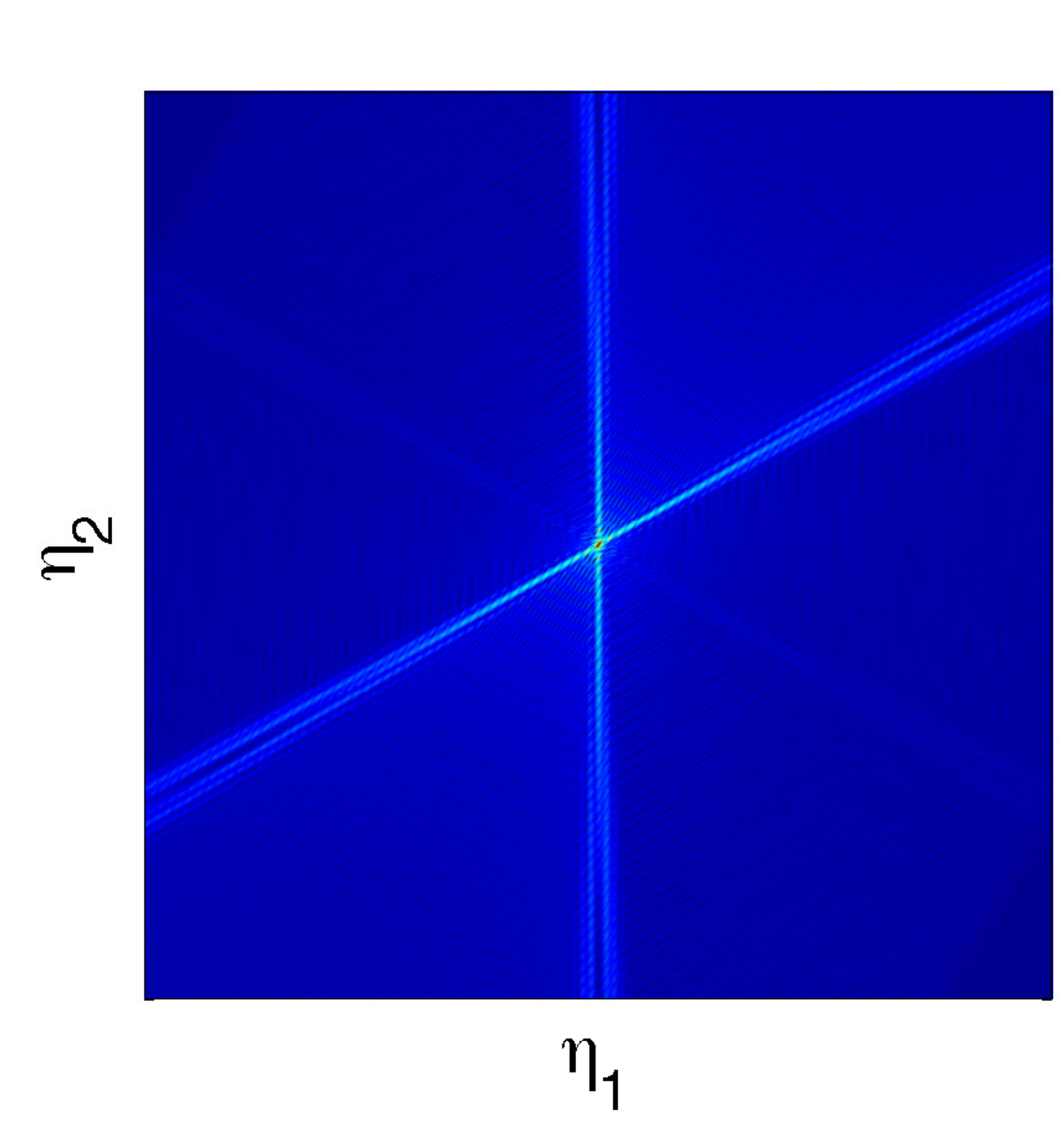}
\caption{\label{fig:hex_resonantZ1}
}
\end{subfigure}
\begin{subfigure}{0.32\linewidth}
\includegraphics[width=\linewidth] {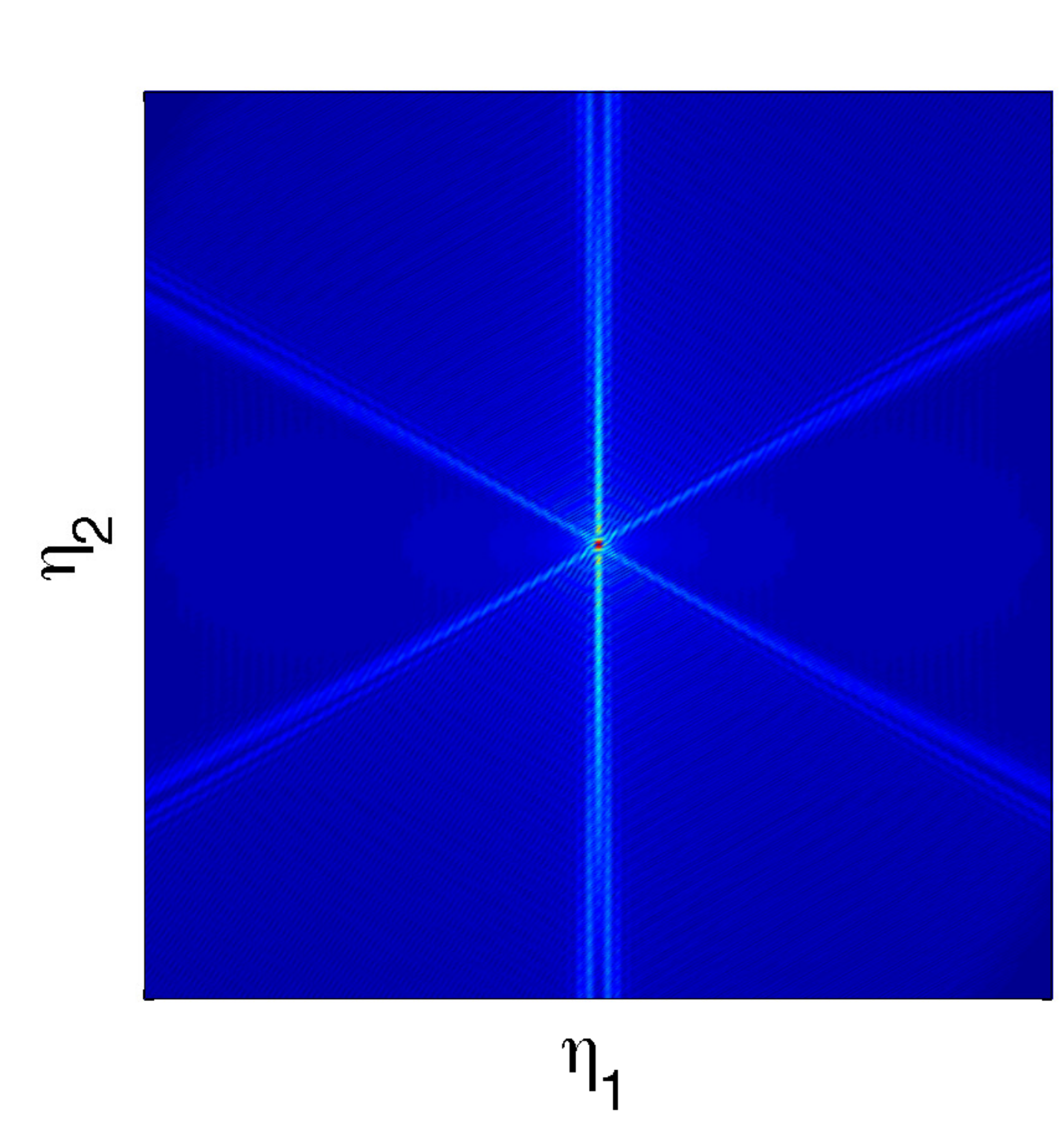}
\caption{\label{fig:hex_resonantZ2}
}
\end{subfigure}
\begin{subfigure}{0.32\linewidth}
\includegraphics[width=\linewidth]{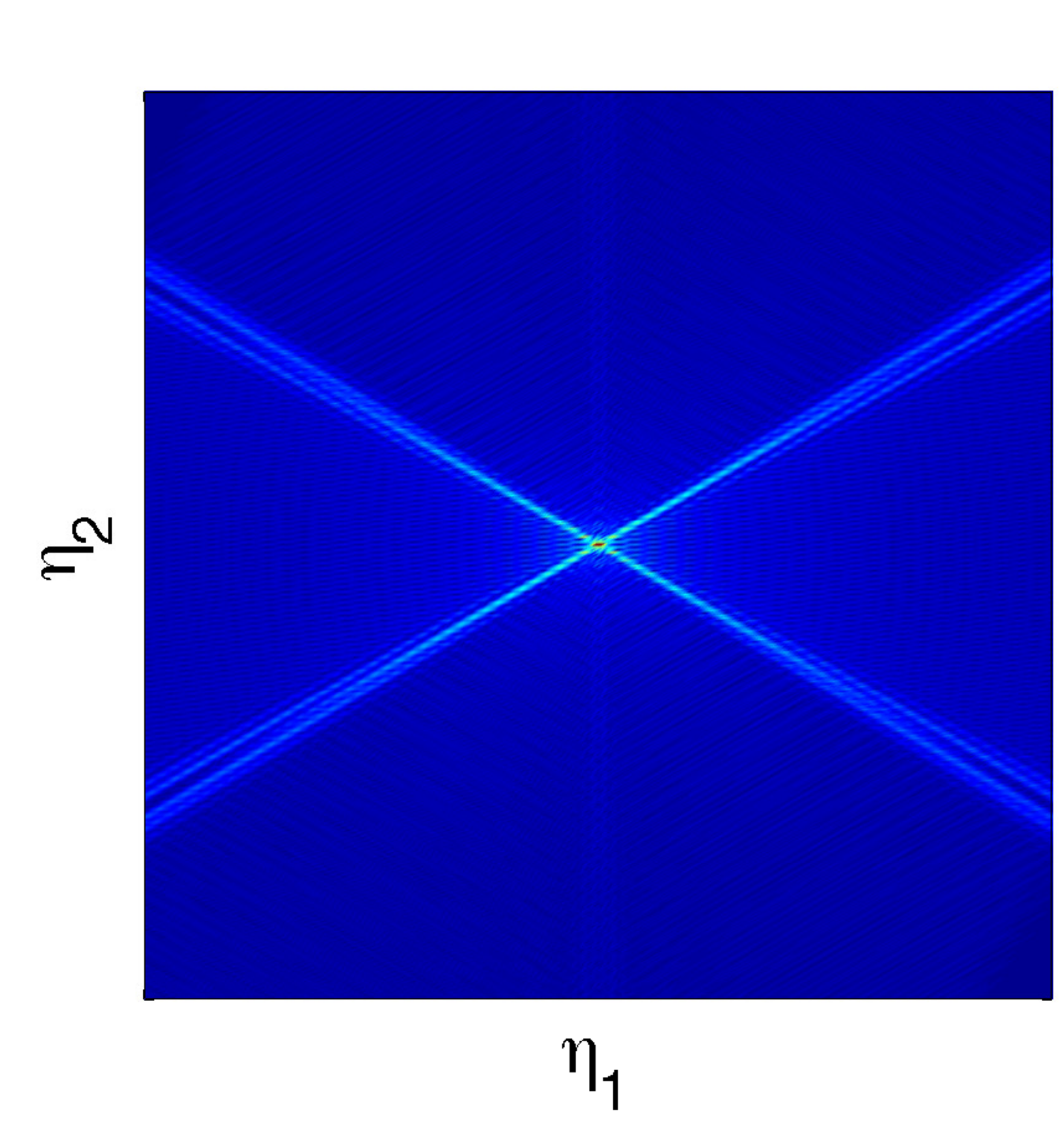}
\caption{\label{fig:hex_resonantZ3}
}
\end{subfigure}
\caption{\label{fig:resonant_freqZ}
The displacement field generated by various forcing configurations at frequencies close to the saddle point frequency at $\omega_0 = 9/4$.
In part (a) the lattice is excited with a forcing vector $\vec{F}=1/2[1,\sqrt{3}]^\mathrm{T}e^{ik_1m_1+im_2(k_1+k_2\sqrt{3})/2}$, where $\vec{k} = \{\pi-\arccos(1/8)\}[1,\sqrt{3}]^\mathrm{T}$, over a cluster consisting of the nodes $\vec{m} \in\{\vec{0},(\pm1,0), (0,\pm1)\}$.
Parts (b) \& (c) show the case when the lattice is forced at node $\vec{m} = \vec{0}$ with forcing vector $\vec{F} = [0,1]^\mathrm{T}$ \& $\vec{F} = [1,0]^\mathrm{T}$ respectively.
In all cases the forcing frequency is $\omega = \sqrt{\omega_0^2 - 0.05}$ and the colour scale is linear from blue (zero) to red (maximal).
}
\end{figure}

At $\gamma$, the leading order problem is 
\begin{equation}
\omega^2_0\vec{u}^{(0)}(\vect{\eta}) = \sigma_0\vec{u}^{(0)}(\vect{\eta}),
\label{eq:leading-order-gamma}
\end{equation}
where $\sigma_0 = -\left[\left(z_1 + z_1^*\right)\mathsf{C}_1 + \left(z_2 + z_2^*\right)\mathsf{C}_2 + \left(z_3 + z_3^*\right)\mathsf{C}_3\right]$ and the phase shifts across the elementary cell are described by $z_1 = -1/8 + i3\sqrt{7}/8$, $z_2 = -31/32 - i3\sqrt{7}/32$, and $z_3 = -1/8 + i3\sqrt{7}/8$, and $(\cdot)^*$ denotes complex conjugation.
The solvability condition gives two eigenfrequencies: $\omega^2 = 27/8$ and $\omega_{(0)}^2 = 81/16$.
The first frequency merely corresponds to the lower dispersion curve at point $\gamma$ in the Brillouin zone, rather than a standing wave frequency, and is therefore not of interest to us.
We instead concentrate on the second eigenmode, which admits the decomposition $\vec{u}^{(0)}(\vect{\eta}) = \phi^{(0)}(\vect{\eta})\vec{U}^{(0)}$, where $\vec{U}^{(0)} = [1/2,\sqrt{3}/2]^\mathrm{T}$.
The solvability of the $\mathcal{O}(\epsilon)$ problem again implies that $\omega_{(1)} = 0$.
The next-to-leading order solution thus admits a solution of the form
\begin{equation}
\vec{u}^{(1)}(\vect{\eta}) = \mathsf{S}_{(0)}^+\sigma_1\vec{U}^{(0)}\phi^{(0)}(\vect{\eta})
+ \left[\mathsf{I} - \mathsf{S}_{(0)}^+\mathsf{S}_{(0)}\right]\vec{v}\psi(\vect{\eta}),
\end{equation}
where $\vec{v}$ and $\psi$ are arbitrary.
The matrix $\mathsf{S}_{(0)}$ is the bracketed term in~\eqref{eq:leading-order-gamma} and
\begin{equation}
\sigma_1 = \left(z_1^* - z_1\right)\mathsf{C}_1\partial_1+ \left(z_2^* - z_2\right)\mathsf{C}_2\left(\frac{1}{2}\partial_1 + \frac{\sqrt{3}}{2}\partial_2\right)
+ \left(z_3^* - z_3\right)\mathsf{C}_3\left(-\frac{1}{2}\partial_1 + \frac{\sqrt{3}}{2}\partial_2\right).
\end{equation}
Moving to the second order problem, we find
\begin{equation}
\omega^2_0\vec{u}^{(2)}(\vect{\eta}) + \omega^2_2\vec{u}^{(0)}(\vect{\eta}) = 
\sigma_{(0)}\vec{u}^{(2)}(\vect{\eta})+\sigma_1\vec{u}^{(1)}(\vect{\eta}) + \sigma_2\vec{u}^{(0)}(\vect{\eta}),
\end{equation}
where
\begin{multline}
\sigma_2 =  -\frac{1}{2}(z_1 + z_1^*)\mathsf{C}_1\partial_1^2
-  \frac{1}{8}(z_2 + z_2^*)\mathsf{C}_2\left(\partial_1^2 + 2\sqrt{3}\partial_1\partial_2 + 3\partial_2^2\right)
\\
-  \frac{1}{8}(z_3 + z_3^*)\mathsf{C}_3\left(\partial_1^2 - 2\sqrt{3}\partial_1\partial_2 + 3\partial_2^2\right).
\end{multline}
Finally, forming the solvability condition as in~\eqref{eq:second-order-pde}, we obtain the following hyperbolic partial differential equation
\begin{equation}
\left(-\frac{45}{64}\partial_1^2 + \frac{9\sqrt{3}}{8}\partial_{12} + \frac{27}{64}\partial_2^2 - \omega_2^2\right)\phi^{(0)}(\vect{\eta}) = 0.
\label{eq:gamma-PDE}
\end{equation}
Since the homogenised equation is hyperbolic, we expect to observe dynamic anisotropy and resonant waveforms when the lattice is excite close to this saddle point frequency.
Figure~\ref{fig:resonant_freqZ} shows the displacement field generated when the triangular lattice is excited near the resonant frequency.
As noted in~\cite{colquitt2012dynamic}, the observed displacement field depends on the orientation of the applied forcing.
For figure~\ref{fig:hex_resonantZ1}, the forcing is applied to a cluster of nodes, in the direction parallel to the inclined bonds (i.e. $[1/2,\sqrt{3}/2]^\mathrm{T}$).
In figures~\ref{fig:hex_resonantZ2} and~\ref{fig:hex_resonantZ3}, the forcing is applied to a single node at $\vec{m} = \vec{0}$ in the vertical and horizontal direction respectively.
In all cases, the directions of preferential propagation are defined by the characteristics of equation~\eqref{eq:gamma-PDE}.
In particular the angles of the preferential directions, as measured counter-clockwise from the $\eta_1$ axis are 
\begin{equation}
\theta_{\pm} = \arctan\left( \pm \frac{3\sqrt{7}}{5} - \frac{4\sqrt{3}}{5} \right),
\end{equation}
or $\theta_+ \approx 0.199$ and $\theta_- \approx 1.25$.
It is remarked that these angles can also be obtained via a quadratic expansion of the dispersion surfaces in the vicinity of the saddle point, as was done in~\cite{movchan2013resonant}.
We also observe that the orientation of these characteristics is different from those which occur at the lower saddle point frequency (cf. equation~\eqref{eq:pointB-homogenised-hyperbolic}).
A similar effect was noted in~\cite{colquitt2012dynamic}.

\subsection{Square lattices}
\label{sec:square}

As mentioned earlier, square lattices are degenerate if the flexural rigidity of the links is neglected.
In contrast to truss-like structures, where the lattice links are connected by pin-joints, the angle at which the links meet is fixed in frame-like structures.
This naturally introduces a new variable: the angle of rotation at the junctions $\theta(\vec{m})$.
In natural units, the equations of motion for the in-plane displacement of a uniform square lattice are
\begin{multline}
\mathsf{M}\omega^2\vec{u}{(\vec{m})} = \mathsf{B}_1\vec{u}(\vec{m}+\vec{e}_1) + \mathsf{B}_2\vec{u}(\vec{m}+\vec{e}_2) + \mathsf{B}_3\vec{u}(\vec{m}-\vec{e}_1) + \mathsf{B}_4\vec{u}(\vec{m}-\vec{e}_2) \\
 + \left[\mathsf{A}_1 + \mathsf{A}_2 + \mathsf{A}_3 + \mathsf{A}_4\right]\vec{u}(\vec{m}),
\end{multline}
where $\mathsf{M} = \diag[1,1,J]$, $J$ is the non-dimensional ratio of mass and polar moment of inertia at the junctions, 
\[
\mathsf{A}_1 = \begin{pmatrix} 1 & 0 & 0 \\ 0 & 6\beta & 3\beta \\ 0 & 3\beta & 2\beta \end{pmatrix},\quad
\mathsf{A}_2 = \begin{pmatrix} 6\beta & 0 & -3\beta \\ 0 & 1 & 0 \\ -3\beta & 0 & 2\beta \end{pmatrix},\quad
\mathsf{A}_3 = \begin{pmatrix} 1 & 0 & 0 \\ 0 & 6\beta & -3\beta \\ 0 & -3\beta & \beta \end{pmatrix},
\]
\[
\mathsf{A}_4 = \begin{pmatrix} 6\beta & 0 & 3\beta \\ 0 & 1 & 0 \\ 3\beta & 0 & 2\beta \end{pmatrix},\quad
\mathsf{B}_1 = \begin{pmatrix} -1 & 0 & 0 \\ 0 & -6\beta & 3\beta \\ 0 & -3\beta & \beta \end{pmatrix},\quad
\mathsf{B}_2 = \begin{pmatrix} -6\beta & 0 & -3\beta \\ 0 & -1 & 0 \\ 3\beta & 0 & \beta \end{pmatrix},
\]
\[
\mathsf{B}_3 = \begin{pmatrix} -1 & 0 & 0 \\ 0 & -6\beta & -3\beta \\ 0 & 3\beta & \beta \end{pmatrix},\quad
\mathsf{B}_4 = \begin{pmatrix} -6\beta & 0 & 3\beta \\ 0 & -1 & 0 \\ -3\beta & 0 & \beta \end{pmatrix},
\]
$\beta = 2D/(\mu\ell^3)$ is the normalised ratio of the flexural rigidity $D$ and longitudinal stiffness $\mu$, and $\vec{u} = [u_1/\ell, u_2/\ell, \theta]^\mathrm{T}$; $\theta$ is the angular rotation.
Here $J = \tilde{J}/(m\ell^2)$, where $\tilde{J}$ is the usual polar moment of inertia.
The frequency is also non-dimensionalised as before.

At this point, it is appropriate to consider the relative size of $\beta$.
The scaled flexural rigidity of the lattice links can also be expressed as $\beta = 2I/(s\ell^2)$, where $I$ is the second moment of inertia and $s$ is the cross-sectional area.
If we assume that the lattice links are solid and have thickness $r$ then $\beta \sim (r/\ell)^2$.
For slender links, as assumed here, $0 < \beta \ll 1$.
As we shall see later, the two free parameters $\beta$ and $J$ give rise to several interesting degeneracies, including Dirac cones.

Once again we proceed with the high frequency homogenisation procedure by introducing a small parameter $0 < \epsilon \ll 1$ and a slow variable $\vect{\eta} = \epsilon\vec{m}$.
The ansatz for the field and frequency are introduced and a hierarchy of equations in ascending orders of $\epsilon$ is obtained, as described in \S\ref{sec:general-theory}.

\begin{figure}
\centering
\begin{minipage}{0.6\linewidth}
\centering
\begin{subfigure}{\linewidth}
\includegraphics[width=\linewidth]{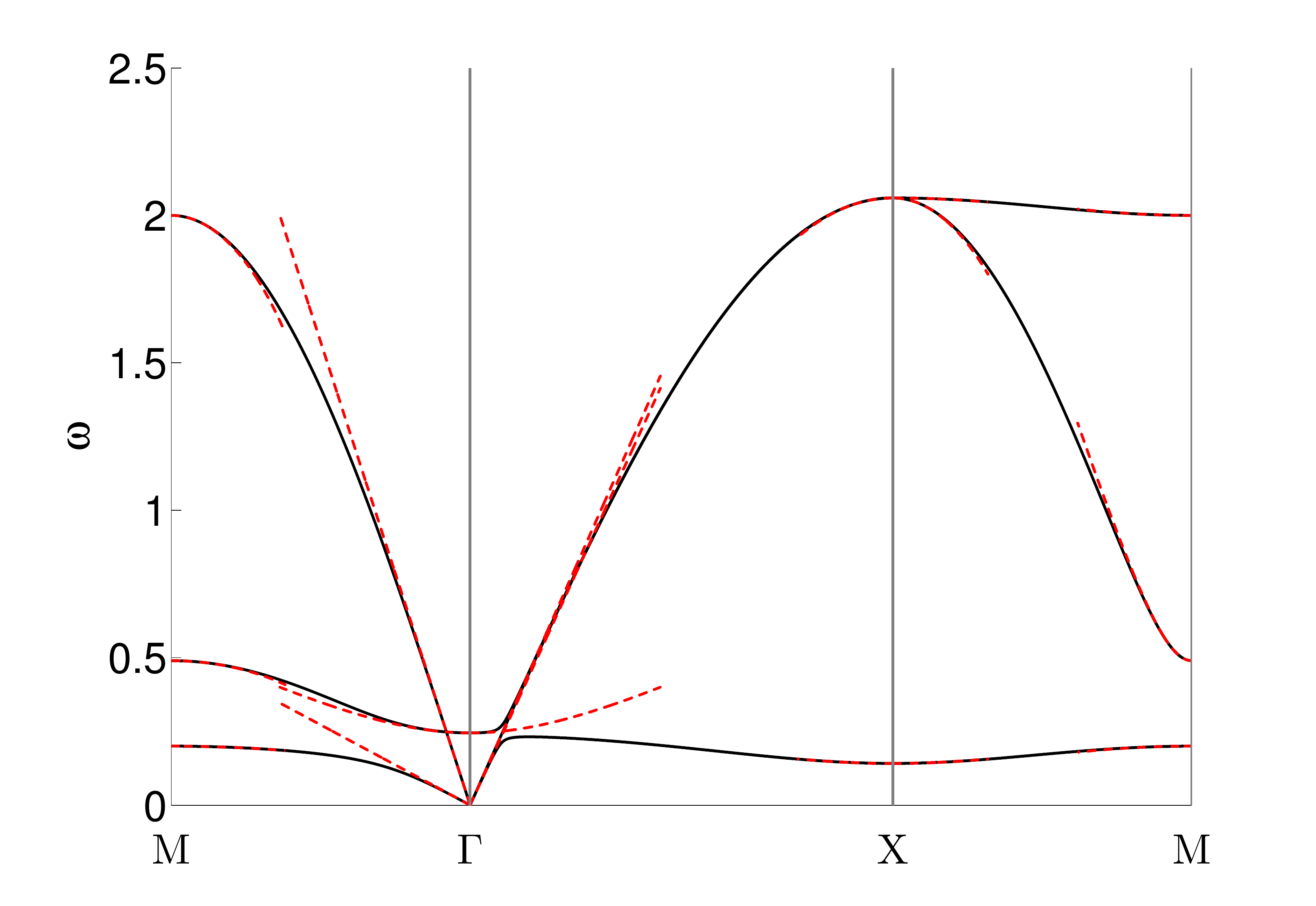}
\caption{\label{square-disp}
$\beta=0.01$, $J=2$
}
\end{subfigure}
\begin{subfigure}{0.45\linewidth}
\includegraphics[width=\linewidth]{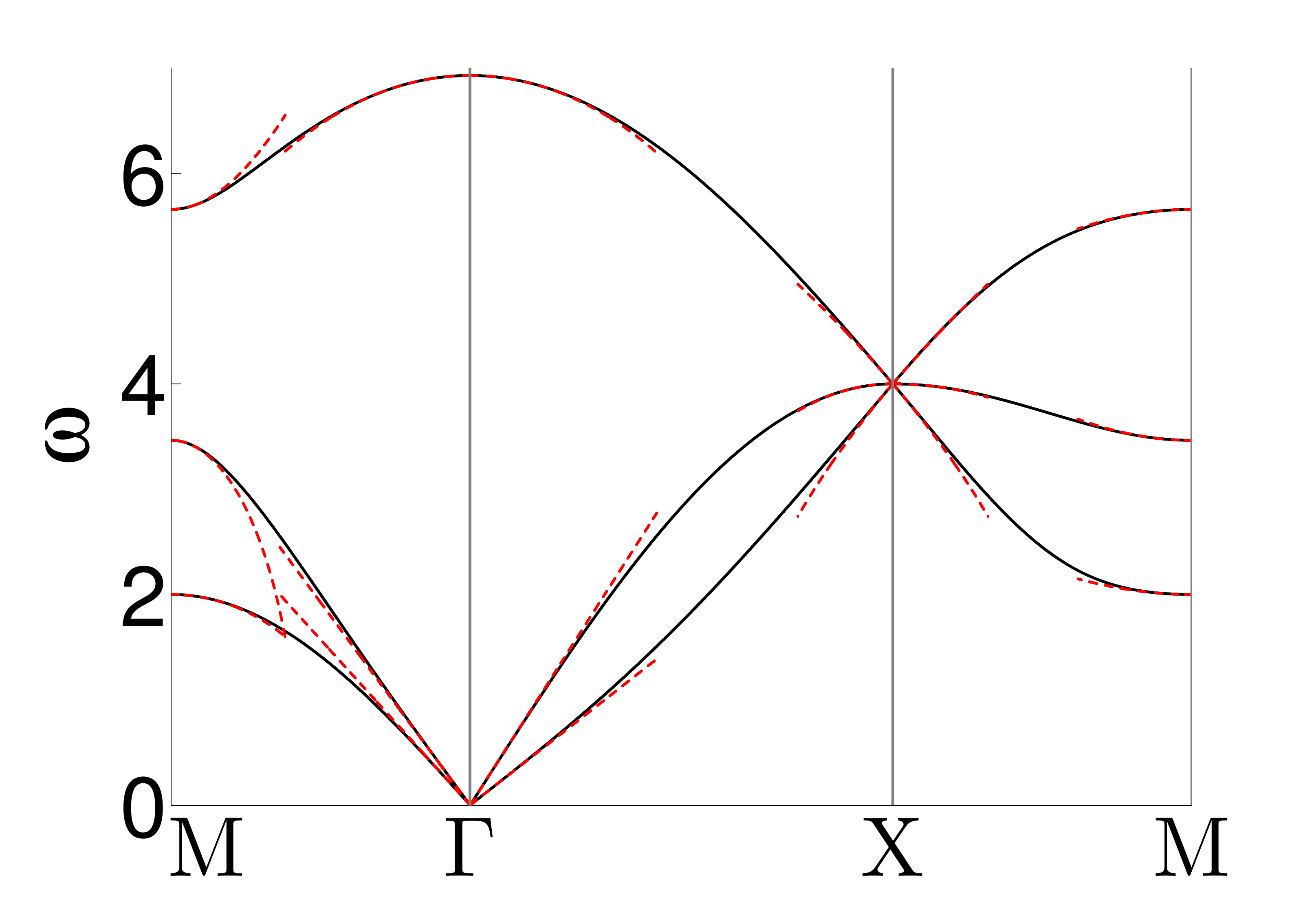}
\caption{\label{square-disp-D1}
$\beta=0.5$, $J=\beta/(1+6\beta)$
}
\end{subfigure}
\begin{subfigure}{0.45\linewidth}
\includegraphics[width=\linewidth]{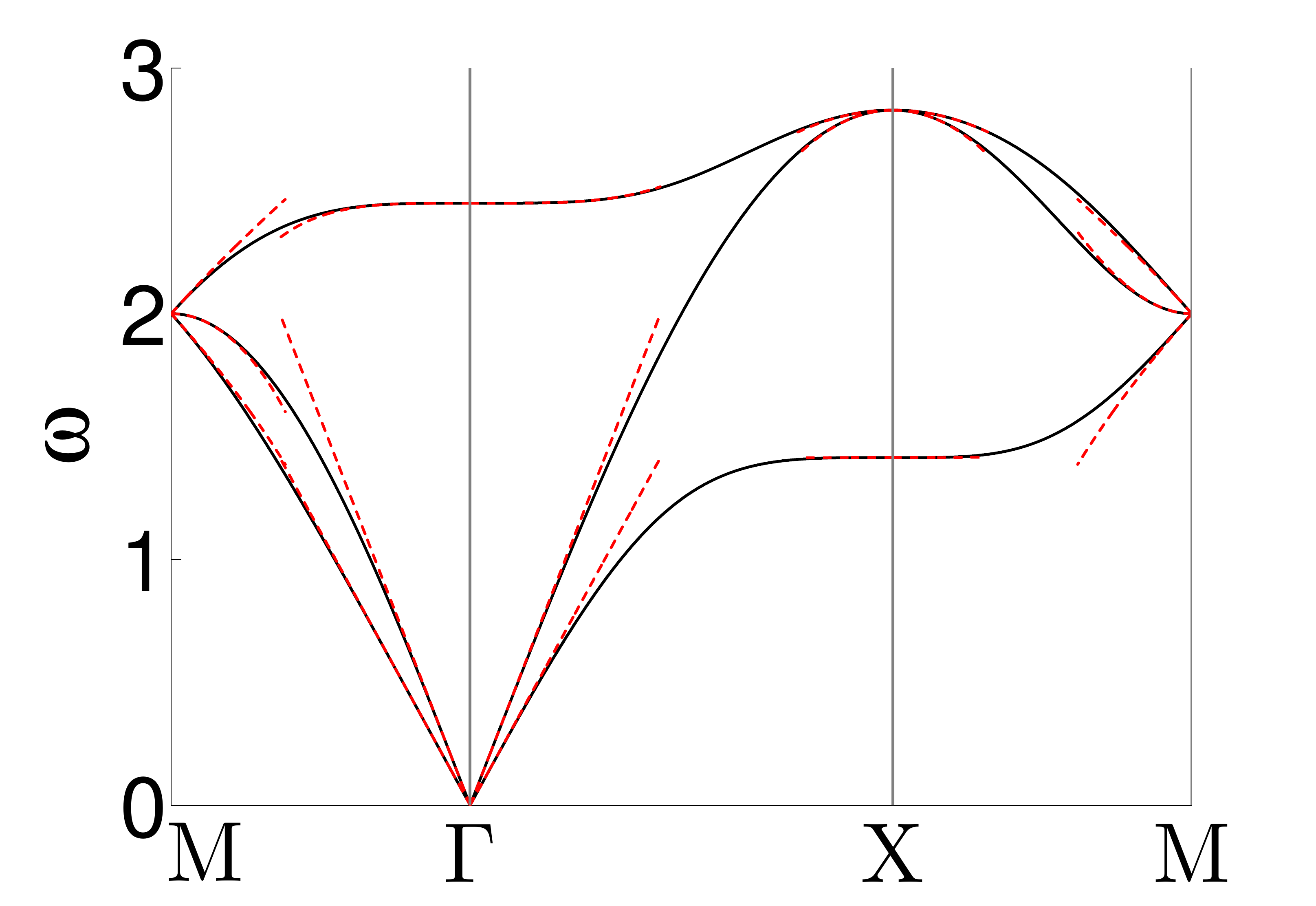}
\caption{\label{square-disp-D2}
$\beta=1/6$, $J=\beta/(1-3\beta)$
}
\end{subfigure}
\end{minipage}
\begin{minipage}{0.27\linewidth}
\begin{subfigure}{\linewidth}
\includegraphics[width=\linewidth]{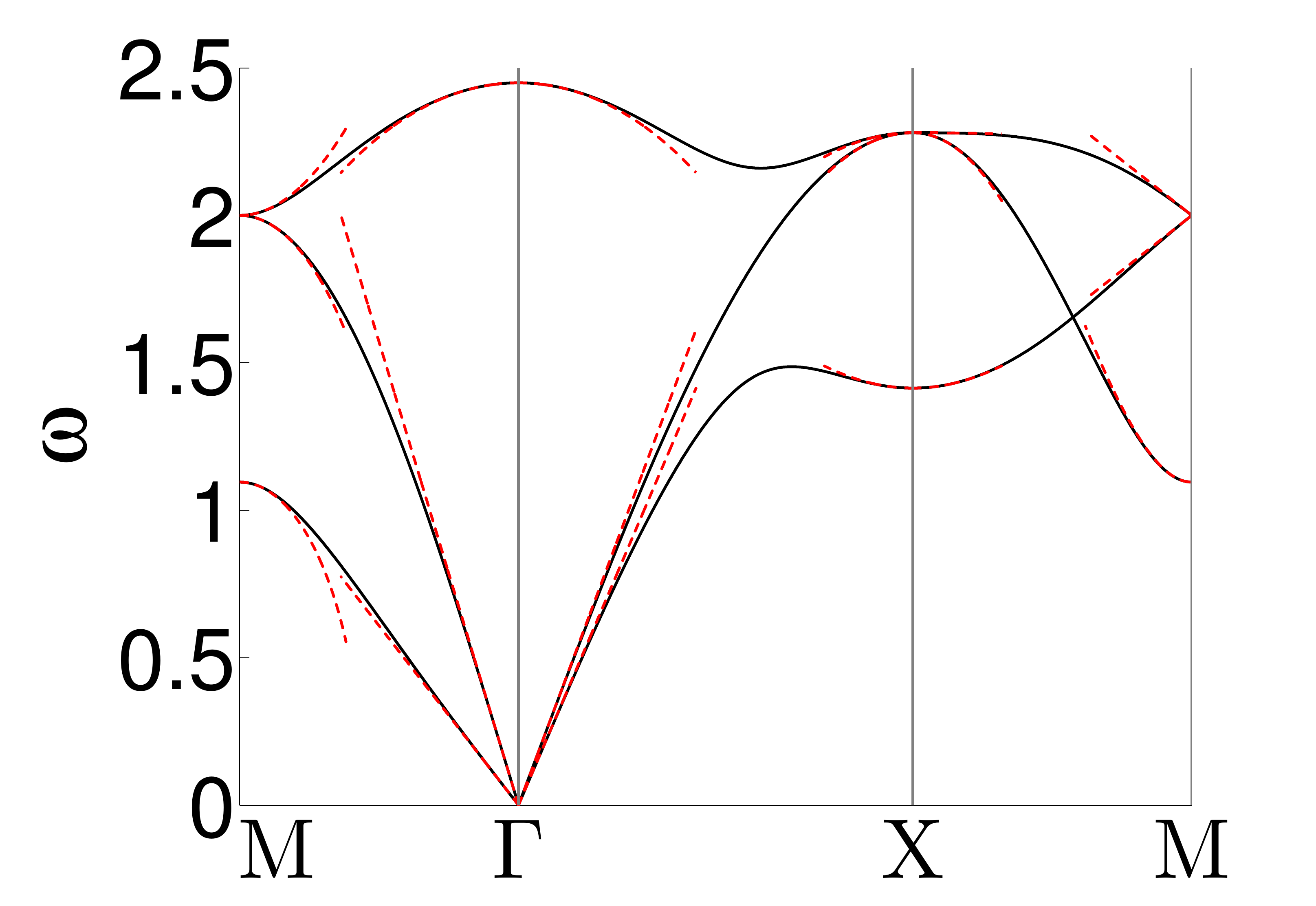}
\caption{\label{square-disp-D3}
$\beta=0.05$, $J=2\beta$
}
\end{subfigure}
\begin{subfigure}{\linewidth}
\includegraphics[width=\linewidth]{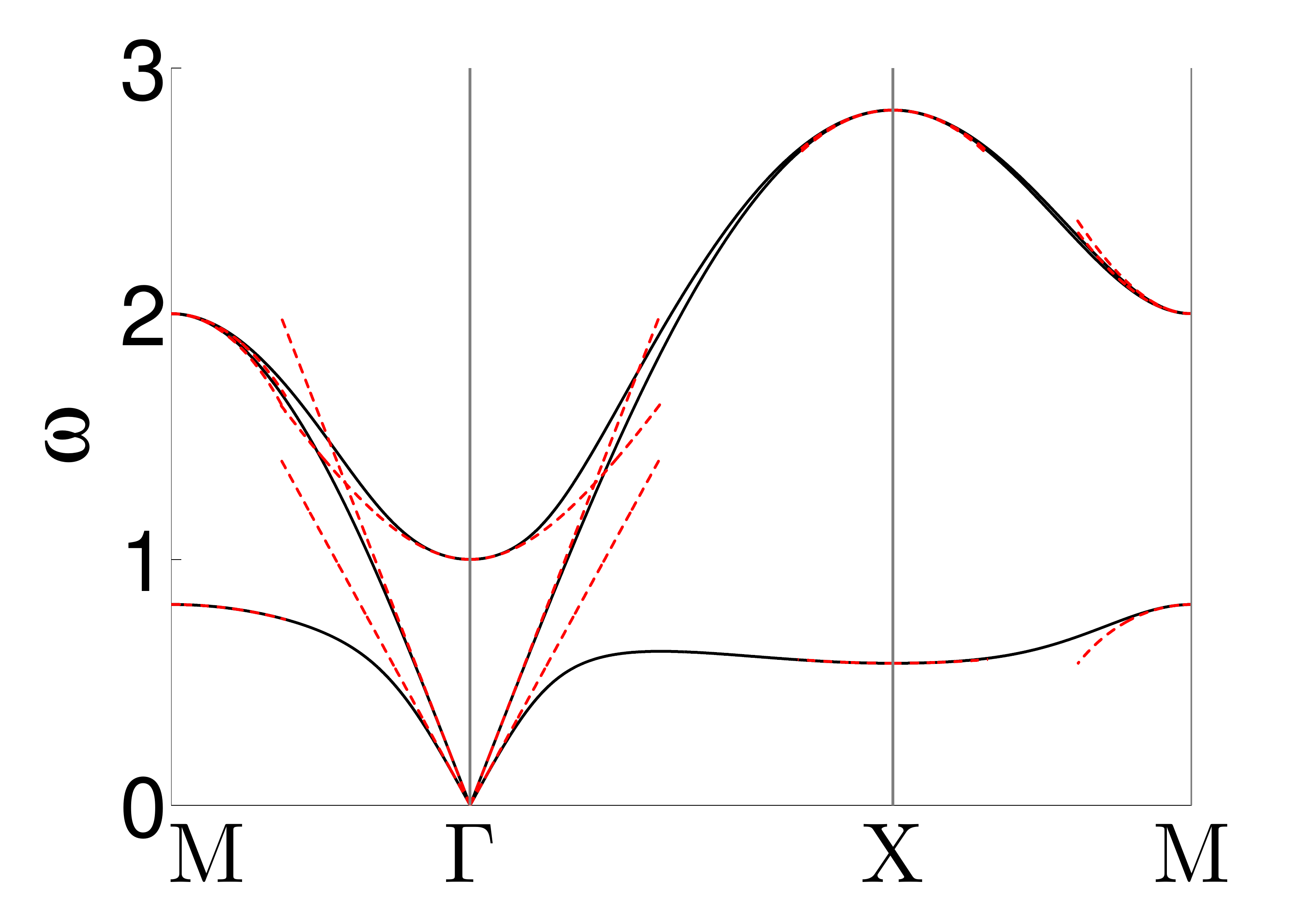}
\caption{\label{square-disp-D4}
$\beta=1/6$, $J=2$
}
\end{subfigure}
\begin{subfigure}{\linewidth}
\includegraphics[width=\linewidth]{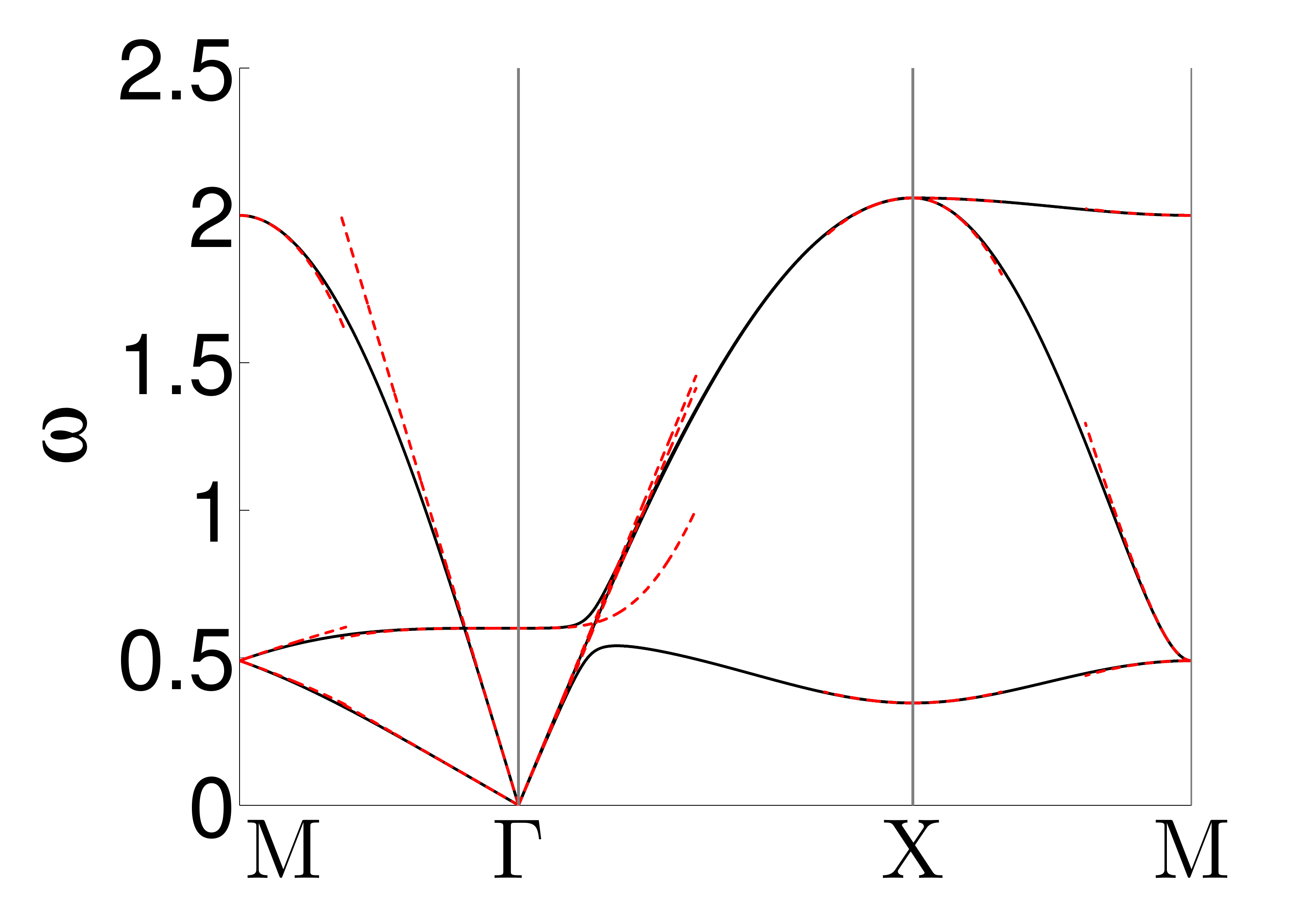}
\caption{\label{square-disp-D5}
$\beta=0.01$, $J=1/3$
}
\end{subfigure}
\end{minipage}
\caption{\label{square-disp-diagrams}
The dispersion diagram for various combinations of the scaled flexural stiffness $\beta$ and moment of inertia $J$.
The dispersion curves are plotted as solid black lines, whilst the asymptotics are shown as dashed red lines.
}
\end{figure}

Figure~\ref{square-disp-diagrams} shows the dispersion diagrams for the square lattice for a range of parameter values.
Figure~\ref{square-disp} shows the dispersion diagram for the case when the parameter values were chosen to be physically reasonable ($\beta = 0.01$ and $J=2$), whereas figures~\ref{square-disp-D1}--\ref{square-disp-D5} are for specific combinations of parameters that correspond to degenerate eigenvalues which will be discussed in the following sections.
It is remarked that, in contrast to the triangular case, the square lattice is not isotropic in the low frequency limit.
Owing to the symmetry of the lattice there are three branches of the dispersion equation corresponding to three different modes of wave propagation.
At standing wave frequencies, these modes can, typically, be decomposed into two translational modes and one rotational mode; away from standing wave frequencies these modes fully couple.

\subsubsection{Point $\mathbf{\Gamma}$: Classical homogenisation}

In this case, at $\vec{k} = \vec{0}$, the leading order problem is
\begin{equation}
\mathsf{M}\omega_0^2\mathsf{u}^{(0)}{(\vect{\eta})} = \left[\mathsf{B}_1+ \mathsf{B}_2 + \mathsf{B}_3  + \mathsf{B}_4
 + \mathsf{A}_1 + \mathsf{A}_2 + \mathsf{A}_3 + \mathsf{A}_4\right]\vec{u}^{(0)}(\vect{\eta}).
\end{equation}
Imposing the usual solvability condition the following eigenvalues are obtained: $\omega_{(0,1)} = 0$ (multiplicity two) and $\omega_{(0,2)} = 2\sqrt{3\beta/J}$.
The leading order solution for the repeated eigenvalue $\omega_0^{(1)} = 0$ then admits the following form
\begin{equation}
\vec{u}^{(0,1)} = \phi^{(0,1,1)}(\vect{\eta})\vec{U}^{(0,1,1)}_1 + \phi^{(0,1,2)}(\vect{\eta})\vec{U}^{(0,1,2)},
\end{equation}
where, without loss of generality, we may take $\vec{U}^{(0,1,i)} = [\delta_{1i},\delta_{2i},0]^\mathrm{T}$.
The distinct eigenvalue admits an associated eigenvector of the form
\begin{equation}
\vec{u}^{(0,2)} = \phi^{(0,2)}(\vect{\eta})\vec{U}^{(0,2)},
\end{equation}
where $\vec{U}^{(0,2)} = [0,0,1]^\mathrm{T}$.
The next to leading order problem yields $\omega_{(i,j)} = 0$ and we move to second order.
For the repeated root $\omega_{(0,1)} = 0$, we obtain a system of two coupled partial differential equations
\begin{subequations}
\begin{equation}
\partial_1^2\phi^{(0,1,1)}(\vect{\eta}) + 3\beta\partial_2^2\phi^{(0,1,1)}(\vect{\eta}) + 3\beta\partial_1\partial_2\phi^{(0,1,2)}(\vect{\eta}) + {\omega_{(2,1)}}^2\phi^{(0,1,1)}(\vect{\eta})  = 0
\end{equation}
and
\begin{equation}
3\beta\partial_1^2\phi^{(0,1,2)}(\vect{\eta}) + \partial_2^2\phi^{(0,1,2)}(\vect{\eta}) + 3\beta\partial_1\partial_2\phi^{(0,1,1)}(\vect{\eta}) + {\omega_{(2,2)}}^2\phi^{(0,1,2)}(\vect{\eta}) = 0.
\end{equation}
\label{eq:square-gamma-coupled}
\end{subequations}
The system decouples into a repeated fourth order partial differential equation for the acoustic modes
\begin{equation}
\left[3\beta\left(\partial_1^4 + \partial_2^4\right) + \partial_1^2\partial_2^2 + m{\omega_{(2,i)}}^2(1+3\beta)\nabla^2 + {\omega_{(2,i)}}^4\right] \phi^{(0,1,i)}(\vect{\eta}) = 0,
\end{equation}
for $i =1,2$.
It is immediately apparent from the homogenised partial differential equation that the square lattice is anisotropic in the low frequency limit.
In the case of a perfect lattice, the asymptotic dispersion equation is linear
\begin{equation}
\omega_{(2,1)}^2\sim \frac{(3\beta+1)}{2m}|\vec{k}|^2 \pm \frac{1}{2m}\sqrt{3\beta(k_1^4+k_2^4)(3\beta-2) + (k_1^2 - k_2^2)^2 + 18k_1^2k_2^2\beta^2 + 12k_1^2k_2^2\beta}.
\end{equation}
The asymptotic dispersion curves are shown as the dashed red lines emanating from the origin in figure~\ref{square-disp-diagrams}.
We observe that the coupled system~\eqref{eq:square-gamma-coupled} is consistent with the homogenised equations derived in~\cite{martinsson2007homogenization} for the static square lattice.

Moving to the simple eigenvalue $\omega_{(0,2)} = 2\sqrt{3\beta/J}$, the first order correction to the frequency again vanishes, and the solvability condition provides the second order partial differential equation governing the leading order envelope function
\begin{equation}
\left(3J - 1\right)\beta\nabla^2\phi^{(0,2)} + J\omega_{(2,2)}^2\phi^{(0,2)} = 0.
\label{eq:gamma-rot1}
\end{equation}

For $J=1/3$, the second order correction to the frequency vanishes and we must proceed to higher order.
Moving to third order we find that $\omega_3$ also vanishes.
Finally, at fourth order we obtain the following partial differential equation for the correction to the frequency and the envelope function
\begin{equation}
\left[\frac{\beta}{4}\left(\partial_1^4 + \partial_2^4\right) - \frac{1}{6}\partial_1^2\partial_2^2 + \omega_4^2\right]\phi^{(0,2)}(\vect{\eta}) = 0.
\label{eq:gamma-rot2}
\end{equation}
Here, the dispersion curve is locally quartic.
Figures~\ref{square-disp}, \ref{square-disp-D1}, \ref{square-disp-D3}. \ref{square-disp-D4} illustrate the quadratic nature of the dispersion curves for this rotational mode near the origin.
In contrast, the quartic behaviour of the curves is illustrated in figures~\ref{square-disp-D2} and~\ref{square-disp-D5}.
Beyond capturing the behaviour of the curves near the origin, the transition of the long scale equation from second order~\eqref{eq:gamma-rot1}, to fourth order~\eqref{eq:gamma-rot2} suggests that, at the critical value of moment of inertia, the lattice transitions from membrane-like behaviour to plate-like behaviour on the long scale.

\subsubsection{At point $\mathbf{M}$: resonant waveforms}
\label{sec:frames-pointB}

The leading order problem is
\begin{equation}
\mathsf{M}\omega_0^2\vec{u}^{(0)}{(\vect{\eta})} = \left[\mathsf{B}_1 - \mathsf{B}_2 + \mathsf{B}_3  - \mathsf{B}_4
 + \mathsf{A}_1 + \mathsf{A}_2 + \mathsf{A}_3 + \mathsf{A}_4\right]\vec{u}^{(0)}(\vect{\eta}).
 \end{equation}
Provided that $J\notin\{1/3, 2\beta\}$ and $\beta\neq1/6$, the solvability criterion yields three simple eigenfrequencies $\omega_{(0,1)} = 2\sqrt{6\beta}$, $\omega_{(0,2)} = 2$, and $\omega_{(0,3)} = 2\sqrt{2\beta/J}$, with associated eigenmodes $\vec{U}^{(0,\ell)} = [\delta_{1\ell},\delta_{2\ell},\delta_{3\ell}]^\mathrm{T}$, where the leading order solution admits the usual decomposition $\vec{u}^{(0,i)}(\vect{\eta}) = \phi^{(0,i)}(\vect{\eta})\vec{U}^{(0,i)}$.
In all cases, the solvability condition for the next to leading order problem imposes $\omega_{(1,i)} = 0$.
Moving to the second order problem, we find a system of three uncoupled partial differential equations for the leading order envelope functions, one for each eigenmode,
\begin{subequations}
\begin{equation}
\left(\partial_1^2 + 3\beta\frac{7-12J}{6J-2}\partial_2^2 + \omega_{(2,1)}^2\right)\phi^{(2,1)}(\vect{\eta}) = 0,
\end{equation}
\begin{equation}
\left(3\beta\frac{2J-\beta}{J-2\beta}\partial_1^2 + \partial_2^2 + \omega_{(2,2)}^2\right)\phi^{(2,2)}(\vect{\eta}) = 0,
\end{equation}
\begin{equation}
\left[\frac{(J+9\beta J - 2\beta)\beta}{J(J-2\beta)}\partial_1^2
+ \frac{(3J+2)\beta}{2J(3J-1)}\partial_2^2
+ \omega_{(2,3)}^2\right]\phi^{(2,3)}(\vect{\eta}) = 0.
\end{equation}
\end{subequations}
The corresponding asymptotics are illustrated by the dashed red curves in figure~\ref{square-disp}.

If $J\in\{1/3, 2\beta\}$ and/or $\beta=1/6$, then one or more of the eigenmodes coincide and the standard decomposition $\vec{u}^{(0,i)}(\vect{\eta}) = \phi^{(0,i)}(\vect{\eta})\vec{U}^{(0,i)}$ no longer applies.
Of particular interest, is the case when $J = 1/3$ and $\beta = 1/6$.
In this case, all three modes collapse down onto a single mode and the leading order eigenvalue, $\omega_0 = 2$, has a multiplicity of three.

\emph{A single degeneracy.}
Let us start by supposing that the first two eigenfrequencies coincide such that $\omega_{(0,1)} = \omega_{(0,2)} = 2$, in which case, the leading order solution is simply a linear combination of the eigenvectors found earlier,
\begin{equation}
\vec{u}^{(0,1)} = \phi^{(0,1,1)}(\vect{\eta})\vec{U}^{(0,1,1)} + \phi^{(0,1,2)}(\vect{\eta})\vec{U}^{(0,1,2)}, 
\end{equation}
where $\vec{U}^{(0,1,i)} = [\delta_{1i},\delta_{2i},0]^\mathrm{T}$.
The third eigenmode remains unchanged.
Imposing the solvability criterion to the first order problem yields $\omega_1 = 0$.
Moving to second order solvability demands
\begin{equation}
0 = {\vec{U}^{(0,1,i)}}^{\dagger}\mathsf{S}_1\mathsf{S}_0^+\mathsf{S}_1{\vec{u}^{(0,1)}} + {\vec{U}^{(0,1,i)}}^{\dagger}\mathsf{S}_2{\vec{u}^{(0,1)}} - \omega_{(0,1)}^2{\vec{U}^{(0,1,i)}}^{\dagger}\mathsf{M}{\vec{u}^{(0,1)}},\;\text{for}\;i=1,2.
\label{eq:B-1st-degen-solvability}
\end{equation}
Equations~\eqref{eq:B-1st-degen-solvability} can be written as a system of two coupled second order partial differential equations, which eventually decouple into a single fourth-order equation
\begin{multline}
\left[ \frac{12J-1}{4(3J-1)}\partial_1^4 + \frac{12J-7}{4(3J-1)}\partial_2^4 - \frac{8-24J}{4(3J-1)}\partial^2_1\partial^2_2
+ \frac{24J-5}{4(3J-1)}\omega_{(2,1)}^2\partial_1^2 \right.\\
\left. - \frac{24J - 11}{4(3J-1)}\omega_{(2,1)}^2\partial_1^2
+ \omega_{(2,1)}^4\right]\phi^{(0,1,i)}(\vect{\eta}) = 0,
\label{eq:sqaure-M-D4}
\end{multline}
for $i=1,2$.
The local dispersion equations for the first two modes are then
\begin{equation}
\omega_{(2,1)}^2 \sim   k_1^2 - k_2^2,\quad \text{and}\quad
\omega_{(2,2)}^2 \sim \frac{12J-1}{12J-4}k_1^2 + \frac{7 - 12J}{12J-4}k_2^2,
\end{equation}
and are illustrated in figure~\ref{square-disp-D4}.
We observe that the homogenised equation~\eqref{eq:sqaure-M-D4} describes an anisotropic medium.
In particular. the curvature of the dispersion curves changes sign at $M$; such effects are associated with dynamic anisotropy as illustrated for triangular lattice in figure~\ref{fig:resonant_freqZ} and discussed in \S\ref{sec:tri-gamma}.

\emph{One sided Dirac cones.}
If second and third eigenmodes coincide, that is, if $\omega_{(0,2)} = \omega_{(0,3)} = 2$ (i.e. $J=2\beta$) the leading order solution admits the decomposition $\vec{u}^{(0,2)} = \phi^{(0,2,1)}(\vect{\eta})\vec{U}^{(0,2,1)} + \phi^{(0,2,2)}(\vect{\eta})\vec{U}^{(0,2,2)}$, with $\vec{U}^{(0,2,i)} = [0,\delta_{1i},\delta_{2i}]^\mathrm{T}$.
Applying the Fredholm alternative to the first order problem leads to a system of first order partial differential equations, which decouples to
\begin{equation}
9J\partial_1^2\phi^{(0,2,i)}(\vect{\eta}) + \omega_{(1,2,i)}^4\phi^{(0,2,i)}(\vect{\eta}) = 0.
\end{equation}
Thus, to leading order, the dispersion curves are linear along contours where $k_1$ is not constant:
\begin{equation}
\omega\sim 2\pm \frac{3\sqrt{J}}{4}k_1,
\end{equation}
suggesting Dirac cone-like behaviour.
If we are interested in behaviour over contours where $k_1$ is constant, then we assume that $\phi^{(0,2,i)}(\vect{\eta}) = \phi^{(0,2,i)}(\eta_2)$ and proceed to higher order.
Imposing the usual solvability condition on the second order problem yields a system of two uncoupled partial differential equations
\begin{subequations}
\begin{equation}
\left(3J\partial_1^2 - \partial_2^2 + \omega_{(2,2,1)}^2\right)\phi^{(0,2,1)} = \partial_1^2v_1(\vect{\eta})
\end{equation}
\begin{equation}
\left[-\frac{1}{2}\partial_1^2 + \frac{1}{2}\left(\frac{9J}{2(3J-1)} - 1\right)\partial_2^2 + \omega_{(2,2,2)}^2\right]\phi^{(0,2,2)} = - 3\partial_1^2v_2(\vect{\eta}),
\end{equation}
\end{subequations}
where $v_i$ are arbitrary functions.
Assuming that the macroscale functions are independent of $\eta_1$ (otherwise the leading order behaviour is linear), we arrive at the equations for the envelope functions
\begin{subequations}
\begin{equation}
\left(\omega_{(2,2,1)}^2  - \partial_2^2\right)\phi^{(0,2,1)} = 0,
\end{equation}
\begin{equation}
\left[\frac{1}{2}\left(\frac{9J}{2(3J-1)} - 1\right)\partial_2^2 + \omega_{(2,2,2)}^2\right]\phi^{(0,2,2)} = 0.
\end{equation}
\end{subequations}
In this case, the dispersion curves are locally quadratic.
This case is illustrated in figure~\ref{square-disp-D3} where we observe that the highest two dispersion curves are linear (Dirac cone like) when approaching $M$ along $XM$, but quadratic when approaching the same point along $\Gamma M$.

If the first and third modes coincide, such that $\omega_{(0,1)} = \omega_{(0,3)} = 2\sqrt{6\beta}$, we arrive at a similar result.
In this case, the degenerate eigenmode admits the decomposition $\vec{u}^{(0,1)}(\vect{\eta}) = \phi^{(0,1,1)}(\vect{\eta})\vec{U}^{(0,1,1)} + \phi^{(0,2,2)}(\vect{\eta})\vec{U}^{(0,1,2)}$ and the leading order behaviour on the long-scale is governed by
\begin{equation}
\left(\frac{36\beta^2}{J} \partial_2^2 + \omega_{(2,1)}^4\right)\phi^{(0,1,i)}(\vect{\eta}) = 0,\quad\text{for}\;i=1,2.
\end{equation}
Moving to higher order and assuming again that the long-scale functions are independent of $\eta_2$, we obtain the following system of uncoupled equations
\begin{subequations}
\begin{equation}
\left[\frac{1}{J}\left(1 + \frac{9\beta^2}{1-6\beta}\right)\partial_1^2 - \omega_{(2,1,1)}\right]\phi^{(0,1,1)} = 0,
\end{equation}
\begin{equation}
\left[\partial_1^2 + \omega_{(2,1,2)}\right]\phi^{(0,1,2)} = 0.
\end{equation}
\end{subequations}
The corresponding asymptotic dispersion curves are shown in figure~\ref{square-disp-D5}.
In this case, the lowest two dispersion curves are linear when approaching $M$ along $\Gamma M$, but quadratic when approaching $M$ along $XM$.

\emph{A triple degeneracy.}
It is interesting to observe what happens when the above two degeneracies coincide, i.e. when $\beta=1/6$ and $J=1/3$.
In this case, we have an eigenvalue $\omega_0 = 2$ of multiplicity three and the solution to the leading order problem admits the form
\begin{equation}
\vec{u}_0(\vect{\eta}) = \left[ \phi^{(1)}(\vect{\eta}), \phi^{(2)}(\vect{\eta}), \phi^{(3)}(\vect{\eta})\right]^\mathrm{T}.
\end{equation}
The next to leading order problem provides a system of three partial differential equations for the unknown scalar functions
\begin{equation}
\vect{0} = \left(\mathsf{S}_1 - \mathsf{M}\omega_1^2\right)\vec{u}_0,
\end{equation}
where
\begin{equation}
\mathsf{S}_1 = \begin{pmatrix}
	0			& 0				& \partial_2 \\
	0			& 0				& \partial _1 \\
	-\partial_2		& -\partial_1		& 0
\end{pmatrix}.
\end{equation}
For non-zero $\omega_1$, the system then decouples into the Helmholtz equation 
\begin{equation}
\left(3\nabla^2 + \omega_1^4\right)\phi^{(i)}(\vect{\eta}) = 0,\quad\text{for}\;i=1,2,3.
\label{eq:M-dirac}
\end{equation}
Thus, the dispersion curves are locally linear and we expect Dirac cone-like behaviour.
In order to obtain the quadratic curve, passing through the Dirac point on the dispersion diagram, we must proceed to higher order.
If $\omega_1$ vanishes, then the first or problem is
\begin{equation}
\mathsf{S}_1\vec{v}(\vect{\eta})  = \begin{bmatrix}
\left(\partial_1^2 - \partial_2^2 + \omega_2^2\right) \phi^{(1)}(\vect{\eta}) \\
\left(\partial_1^2 - \partial_2^2 + \omega_2^2\right) \phi^{(2)}(\vect{\eta}) \\
\frac{1}{6}\left(-\partial_1^2 + \partial_2^2 + 2\omega_2^2\right) \phi^{(3)}(\vect{\eta})
\end{bmatrix},
\end{equation}
where $\vec{v}(\vect{\eta})$ is an arbitrary vector function of $\vect{\eta}$.
For vanishing $\omega_1$, the first order problem requires that $\partial_1\phi^{(3)}  = \partial_2\phi^{(3)}  = 0$, whence $\phi^{(3)} = c_1$.
Additionally, $\partial_1\phi^{(2)} + \partial_2\phi^{(1)} = 0$ and hence $\phi^{(1)} = \partial_1\mathcal{H}(\vect{\eta})$ and $\phi^{(2)} = -\partial_2\mathcal{H}(\vect{\eta})$.
Thus, for the case when the structure is locally invariant with respect to $\eta_2$, $\phi^{(2)} = 0$ and we obtain a single differential equation for the envelope function
\begin{subequations}
\begin{equation}
\left(\frac{d^2}{d\eta_1^2} + \omega_2^2\right)\phi^{(1)}(\eta_1) = 0,
\label{eq:M-dirac-elliptic}
\end{equation}
similarly, when the system in locally independent of $\eta_1$, we find $\phi^{(1)} = 0$, and
\begin{equation}
\left(\frac{d^2}{d\eta_2^2} - \omega_2^2\right)\phi^{(2)}(\eta_2) = 0.
\label{eq:M-dirac-hyperbolic}
\end{equation}
\end{subequations}
Hence, we obtain the quadratic dispersion curves passing through the Dirac point on the dispersion diagram~\ref{square-disp-D2}.
We observe from the homogenised equation~\eqref{eq:M-dirac}, that the Dirac cone has circular cross section and, hence, the response of the lattice will be isotropic.
We also observe that the homogenised partial differential equations which govern the long scale behaviour of the third mode transition from elliptic~\eqref{eq:M-dirac-elliptic} to hyperbolic~\eqref{eq:M-dirac-hyperbolic} as we move through point $M$ (see figure~\ref{square-disp-D2}).
As discussed earlier, such behaviour is associated with dynamic anisotropy and wave beaming.
In addition to those degeneracies already mentioned, the envelope functions become independent of one of the slow variables when $J = 7/12$ and $\beta = 2J$.

\subsubsection{At point $\mathbf{X}$}

In this case neighbouring nodes oscillate out of phase with each other and the leading order problem is
\begin{equation}
\mathsf{M}\omega_0^2\vec{u}^{(0)}{(\vect{\eta})} = \left[\mathsf{A}_1 + \mathsf{A}_2 + \mathsf{A}_3 + \mathsf{A}_4 - \mathsf{B}_1- \mathsf{B}_2 - \mathsf{B}_3  - \mathsf{B}_4\right]\vec{u}^{(0)}(\vect{\eta}).
\end{equation}
Provided that $J \neq \beta/(1+6\beta)$, we obtain two eigenvalues, the first of which is $\omega_{(0,1)} = 2\sqrt{1+6\beta}$ and has multiplicity two; the second is $\omega_{(0,2)} = 2\sqrt{\beta/J}$ and has unit multiplicity.
The leading order eigenmodes then admit the following representation
\begin{equation}
\vec{u}^{(0,1)} = \phi^{(0,1,1)}(\vect{\eta})\vec{U}^{(0,1,1)} + \phi^{(0,1,2)}(\vect{\eta})\vec{U}^{(0,1,2)},\;\text{and}\;
\vec{u}^{(0,2)} = \phi^{(0,2)}(\vect{\eta})\vec{U}^{(0,2)},
\end{equation}
where $\vec{U}^{(0,1,i)} = [\delta_{1i},\delta_{2i},0]^\mathrm{T}$ and $\vec{U}^{(0,2)} = [0,0,1]^\mathrm{T}$.
It is observed that, once again, the translational and rotational modes decouple at leading order.
At next to leading order, the solvability criterion requires that $\omega_{(1,i)} = 0$.
For the first eigenmode, the solvability criterion yields a system of second order partial differential equations, which eventually decouple as
\begin{multline}
\beta\left[ 6J + \beta(36J-15)\right]\left[\partial_1^4 + \partial_2^4\right]
+\left[(216J - 144)\beta^3 + 36\beta^2J + (6J-m)\beta + J\right]\partial_1^2\partial_2^2 \\
-36\omega_{(2,i)}^2\left\{\left[\left(J - \frac{5}{12}\right)\beta^2 + \left(\frac{J}{3} - \frac{1}{36}\right)\beta + \frac{J}{36}\right]\left[\partial_1^2 + \partial_2^2\right] \right.\\
\left. - \frac{J}{6}\omega_{(2,i)}^2\left[\left(J - \frac{1}{6}\right)\beta + \frac{J}{6}\right]\right\}\phi^{(0,i)}(\vect{\eta}) = 0.
\end{multline}
For the second, simple eigenmode, we find a single second order partial differential equation for the envelope function
\begin{equation}
\left[\left(\frac{9J\beta^2}{\beta - J(1+6\beta)}+\beta\right)\left(\partial_1^2 + \partial_2^2\right) + J\omega_{(2,2)}^2\right]\phi^{(0,2)}(\vect{\eta}) = 0.
\label{eq:X-2nd-order}
\end{equation}
For $J\neq\beta/(1-3\beta)$, the above equation relates the correction to the frequency and the leading order envelope function.
If however, $J=\beta/(1-3\beta)$, then $\omega_{(2,2)}=0$ and we must proceed to higher order where we find that $\omega_{(3,2)}$ also vanishes.
Finally, at fourth order we obtain a single partial differential equation linking the leading order envelope function $\phi^{(0,2)}(\vect{\eta})$ and the correction to the frequency $\omega_{(4,2)}$
\begin{equation}
\left[ \frac{(1-3\beta)\beta}{12} \partial_1^4 + \frac{\beta(1-3\beta)}{12} \partial_2^4 -\frac{(1-3\beta)}{18} \partial_1^2\partial_2^2 + \omega_{(4,2)}^2\right]\phi^{(0,2)}(\vect{\eta}) = 0.
\label{eq:X-4th-order}
\end{equation}
The corresponding asymptotic dispersion curves are shown in figure~\ref{square-disp-D2}, when the dispersion curve for the lowest mode is almost flat.
Once again, for a critical value of $J$, we observe a transition in the homogenised partial differential equation, from second~\eqref{eq:X-2nd-order} to fourth order~\eqref{eq:X-4th-order}.

At point $X$, the three modes coincide when $J = \beta/(1+6\beta)$ such that we obtain a single eigenvalue $\omega_{(0)} = 2\sqrt{1+6\beta}$ of multiplicity three.
We then proceed as in the triply degenerate case discussed in \S\ref{sec:frames-pointB} and obtain a system of three coupled partial differential equations.
For non-vanishing $\omega_{(1)}$, the system can be decoupled to obtain three identical uncoupled equations
\begin{equation}
\left\{36\beta(1+36\beta)\left[\partial_1^2 + \partial_2^2\right] + \omega_{(1)}^4\right\}\phi^{(0,i)}(\vect{\eta}) = 0,
\qquad\text{for}\;
i=1,2,3,
\end{equation}
and we obtain the two linear dispersion curves characteristic of Dirac cones (cf. figure~\ref{square-disp-D1}).
As before, in order to obtain the quadratic curve bisecting the Dirac cone, we must proceed to higher order.
In doing so, we obtain the following two dispersion curves about point $X$
\begin{subequations}
\begin{equation}
\omega^2 \sim 4(1+6\beta) - (1+6\beta)|\vec{k}|^2,
\end{equation}
along $X\Gamma$ and
\begin{equation}
\omega^2 \sim 4(1+6\beta) - k_1^2,
\end{equation}
along $XM$.
\end{subequations}

It is interesting to observe that, in contrast to the Dirac cone for the triangular lattice discussed in \S\ref{sec:triangular-X}, this Dirac cone is created by a degeneracy and only exists for a certain combination of material parameters.

\section{Concluding remarks}
\label{sec:conclusion}

The asymptotic theory developed herein can be applied to a wide range of discrete structures of arbitrary geometry and dimension.
Indeed, any discrete system where the interaction (e.g. equilibrium equation) between points is linear can be analysed using the scheme presented herein.
As outlined in the introduction, lattice type structures are of significant interest in a wide range of physical settings, including biomechanics, structural mechanics, and cloaking.
Often, one can obtain effective material properties from analysing the static response of the lattice; but such approaches are limited to the low-frequency regime whereas many novel features associated with metamaterials, such as cloaking, dynamic anisotropy and focusing occur at higher frequencies.
The two-scale approach used here allows us to obtain effective material properties in the vicinity of any standing wave frequency.
A detailed understanding of the material response at higher frequencies would, potentially, allow the design of metamaterial devices which are effective over a much wider range of frequencies.

The general theoretical methodology is accompanied by illustrative examples for two archetypal two-dimensional lattices: triangular in \S\ref{sec:triangular} and square in \S\ref{sec:square}.
As demonstrated in figure~\ref{fig:resonant_freqZ} in \S\ref{sec:tri-gamma}, the two scale approach used here accurately captures the essential dynamic behaviour at resonant frequencies away from the low-frequency regime.
Moreover, the high frequency homogenisation methodology also captures the interesting behaviour associated with degeneracies in discrete systems (cf. figure~\ref{square-disp-diagrams}).
In particular, where such degeneracies occur, the asymptotic procedure elucidates changes in the governing equations of the system on the long scale leading to physical insight; for example, moving from membrane-like to plate-like behaviour (cf. equations~\eqref{eq:gamma-rot2} and~\eqref{eq:gamma-rot2} on p.~\pageref{eq:gamma-rot1}).

\section*{Acknowledgements}
DJC gratefully acknowledges financial support from the EPSRC in the form of a Doctoral Prize Fellowship and grant EP/J009636/1.
RVC thanks the EPSRC for their support through research grants EP/I018948/1, EP/L024926/1, EP/J009636/1 and Mathematics Platform grant EP/I019111/1. 

\bibliographystyle{qjmam}
\bibliography{HFH-Lattices-Refs}

\end{document}

%% file: uniform-triangular-lattice.tex
\begin{tikzpicture}[scale=1]

\foreach \i in {-2,-1,0,1,2} {
	\draw[line width=1.25, color=gray] (\i-0.5,0) -- (\i+0.5,0);
	\draw[line width=1.25, color=gray, rotate around={60:(\i,0)}] (\i-0.5,0) -- (\i+0.5,0);
	\draw[line width=1.25, color=gray, rotate around={120:(\i,0)}] (\i-0.5,0) -- (\i+0.5,0);
	\draw[shading=ball, ball color=black] (\i,0) circle (0.15);
}

\begin{scope}[shift={(0.5,0.5*3^(0.5))}]
\foreach \i in {-2,-1,0,1,2} {
	\draw[line width=1.25, color=gray] (\i-0.5,0) -- (\i+0.5,0);
	\draw[line width=1.25, color=gray, rotate around={60:(\i,0)}] (\i-0.5,0) -- (\i+0.5,0);
	\draw[line width=1.25, color=gray, rotate around={120:(\i,0)}] (\i-0.5,0) -- (\i+0.5,0);
	\draw[shading=ball, ball color=black] (\i,0) circle (0.15);
}

\end{scope}

\begin{scope}[shift={(0,3^(0.5))}]
\foreach \i in {-2,-1,0,1,2} {
	\draw[line width=1.25, color=gray] (\i-0.5,0) -- (\i+0.5,0);
	\draw[line width=1.25, color=gray, rotate around={60:(\i,0)}] (\i-0.5,0) -- (\i+0.5,0);
	\draw[line width=1.25, color=gray, rotate around={120:(\i,0)}] (\i-0.5,0) -- (\i+0.5,0);
	\draw[shading=ball, ball color=black] (\i,0) circle (0.15);
}

\end{scope}

\begin{scope}[shift={(0.5,-0.5*3^(0.5))}]
\foreach \i in {-2,-1,0,1,2} {
	\draw[line width=1.25, color=gray] (\i-0.5,0) -- (\i+0.5,0);
	\draw[line width=1.25, color=gray, rotate around={60:(\i,0)}] (\i-0.5,0) -- (\i+0.5,0);
	\draw[line width=1.25, color=gray, rotate around={120:(\i,0)}] (\i-0.5,0) -- (\i+0.5,0);
	\draw[shading=ball, ball color=black] (\i,0) circle (0.15);
}

\end{scope}

\begin{scope}[shift={(0,-3^(0.5))}]
\foreach \i in {-2,-1,0,1,2} {
	\draw[line width=1.25, color=gray] (\i-0.5,0) -- (\i+0.5,0);
	\draw[line width=1.25, color=gray, rotate around={60:(\i,0)}] (\i-0.5,0) -- (\i+0.5,0);
	\draw[line width=1.25, color=gray, rotate around={120:(\i,0)}] (\i-0.5,0) -- (\i+0.5,0);
	\draw[shading=ball, ball color=black] (\i,0) circle (0.15);
}
\end{scope}

\end{tikzpicture}

%% file: uniform-square-lattice.tex
\begin{tikzpicture}[scale=1]
\clip (-2.5,-2.5) rectangle (2.5,2.5);

\draw[line width=1.25, color=gray] (-3,0) -- (3,0);
\draw[line width=1.25, color=gray]  (-3,1) -- (3,1);
\draw[line width=1.25, color=gray]  (-3,2) -- (3,2);
\draw[line width=1.25, color=gray]  (-3,-1) -- (3,-1);
\draw[line width=1.25, color=gray]  (-3,-2) -- (3,-2);

\draw[line width=1.25, color=gray]  (0,-3) -- (0,3);
\draw[line width=1.25, color=gray]  (-1,-3) -- (-1,3);
\draw[line width=1.25, color=gray]  (-2,-3) -- (-2,3);
\draw[line width=1.25, color=gray]  (1,-3) -- (1,3);
\draw[line width=1.25, color=gray]  (2,-3) -- (2,3);

\draw[shading=ball, ball color=black] (-2,-2) circle (0.15);
\draw[shading=ball, ball color=black] (-1,-2) circle (0.15);
\draw[shading=ball, ball color=black] (0,-2) circle (0.15);
\draw[shading=ball, ball color=black] (1,-2) circle (0.15);
\draw[shading=ball, ball color=black] (2,-2) circle (0.15);

\draw[shading=ball, ball color=black] (-2,-1) circle (0.15);
\draw[shading=ball, ball color=black] (-1,-1) circle (0.15);
\draw[shading=ball, ball color=black] (0,-1) circle (0.15);
\draw[shading=ball, ball color=black] (1,-1) circle (0.15);
\draw[shading=ball, ball color=black] (2,-1) circle (0.15);

\draw[shading=ball, ball color=black] (-2,0) circle (0.15);
\draw[shading=ball, ball color=black] (-1,0) circle (0.15);
\draw[shading=ball, ball color=black] (0,0) circle (0.15);
\draw[shading=ball, ball color=black] (1,0) circle (0.15);
\draw[shading=ball, ball color=black] (2,0) circle (0.15);

\draw[shading=ball, ball color=black] (-2,2) circle (0.15);
\draw[shading=ball, ball color=black] (-1,2) circle (0.15);
\draw[shading=ball, ball color=black] (0,2) circle (0.15);
\draw[shading=ball, ball color=black] (1,2) circle (0.15);
\draw[shading=ball, ball color=black] (2,2) circle (0.15);

\draw[shading=ball, ball color=black] (-2,1) circle (0.15);
\draw[shading=ball, ball color=black] (-1,1) circle (0.15);
\draw[shading=ball, ball color=black] (0,1) circle (0.15);
\draw[shading=ball, ball color=black] (1,1) circle (0.15);
\draw[shading=ball, ball color=black] (2,1) circle (0.15);

\end{tikzpicture}